\documentclass[superscriptaddress,letterpaper,aps,twocolumn]{revtex4-1}

\usepackage[dvips]{graphicx}
\usepackage[T1]{fontenc}
\usepackage{amssymb,amsfonts,amsmath}

\newcommand{\mb}[1]{\mathbf{#1}}

\begin{document}

\title{Deconvolution of dynamic mechanical networks}

\author{Michael Hinczewski}
\affiliation{Institute for Physical Science and Technology, University of Maryland, College Park, MD 20742}
\email{mhincz@umd.edu}
\affiliation{Department of Physics, Technical University of Munich, 85748 Garching, Germany}

\author{Yann von Hansen}
\author{Roland R. Netz}
\affiliation{Department of Physics, Technical University of Munich, 85748 Garching, Germany}

\begin{abstract}
Time-resolved single-molecule biophysical experiments yield data that
contain a wealth of dynamic information, in addition to the
equilibrium distributions derived from histograms of the time
series. In typical force spectroscopic setups the molecule is
connected via linkers to a read-out device, forming a mechanically
coupled dynamic network.  Deconvolution of equilibrium distributions,
filtering out the influence of the linkers, is a straightforward and
common practice. We have developed an analogous dynamic deconvolution
theory for the more challenging task of extracting kinetic properties
of individual components in networks of arbitrary complexity and
topology.  Our method determines the intrinsic linear response
functions of a given molecule in the network, describing the power
spectrum of conformational fluctuations.  The practicality of our
approach is demonstrated for the particular case of a protein linked
via DNA handles to two optically trapped beads at constant stretching
force, which we mimic through Brownian dynamics simulations.  Each
well in the protein free energy landscape (corresponding to folded,
unfolded, or possibly intermediate states) will have its own
characteristic equilibrium fluctuations.  The associated linear
response function is rich in physical content, since it depends both
on the shape of the well and its diffusivity---a measure of the
internal friction arising from such processes like the transient
breaking and reformation of bonds in the protein structure.  Starting
from the autocorrelation functions of the equilibrium bead
fluctuations measured in this force clamp setup, we show how an
experimentalist can accurately extract the state-dependent protein
diffusivity using a straightforward two-step procedure.
\end{abstract}

\maketitle

Force spectroscopy of single biomolecules relies most
commonly on atomic force microscope
(AFM)~\cite{CarrionVazquez99,Oesterhelt00,Wiita07,Khatri07,Greene08,Puchner08,Junker09}
or optical
tweezer~\cite{Cecconi05,Woodside06,Woodside06b,Greenleaf08,Chen09,Gebhardt10}
techniques.  By recording distance fluctuations under applied tension,
these experiments serve as sensitive probes of free energy
landscapes~\cite{Cecconi05,Woodside06,Woodside06b,Greenleaf08,Gebhardt10},
and structural transformations associated with ligand binding or
enzymatic activity~\cite{Wiita07,Puchner08,Junker09}.  All such
studies share an unavoidable complication: the signal of interest is
the molecule extension as a function of time, but the experimental
output signal is an indirect measure like the deflection of the AFM
cantilever or the positions of beads in optical traps.  The signal is
distorted through all elements in the system, which in addition
typically include polymeric handles such as protein domains or
double-stranded DNA which connect the biomolecule to the cantilever or
bead.  As shown in the case of an RNA hairpin in an optical
tweezer~\cite{Hyeon08}, handle fluctuations lead to nontrivial
distortions in equilibrium properties like the energy landscape as
well as dynamic quantities like folding/unfolding rates.  If an
accurate estimate of the biomolecule properties is the goal, then one
needs a systematic procedure to subtract the extraneous effects and
recover the original signal from experimental time series data.

{\it Static deconvolution}, which operates on the equilibrium
distribution functions of connected objects, is a well-known
statistical mechanics procedure and has been successfully applied to
recover the free energy landscape of DNA
hairpins~\cite{Woodside06,Woodside06b} and more recently of a leucine
zipper domain~\cite{Gebhardt10}.  In contrast, for dynamic properties
of the biomolecule, no comprehensive deconvolution method exists.
Handles and beads have their own dissipative characteristics and will
tend to suppress the high-frequency fluctuations of the biomolecule
and as a result distort the measured power spectrum.  In the context
of single-molecule pulling experiments, theoretical progress has been
made in accounting for handle/bead effects on the observed unfolding
transition rates~\cite{Manosas2005,Dudko2006,Manosas2007,Dudko2008}.
However, the full intrinsic fluctuation spectrum, as encoded in the
time-dependent linear response function, has remained out of reach.
The current work presents a systematic {\it dynamic deconvolution}
procedure that fills this gap, providing a way to recover the linear
response of a biomolecule integrated into a mechanical dissipative
network.  We work in the constant force ensemble as appropriate for
optical force clamp setups with active feedback
mechanisms~\cite{Lang02,Nambiar04} or passive
means~\cite{Greenleaf05,Chen09}.  While our theory is general and
applies to mechanical networks of arbitrary topology, we illustrate
our approach for the specific experimental setup of
Ref.~\cite{Gebhardt10}: a protein attached to optically trapped beads
through dsDNA handles.  The only inputs required by our theory
  are the autocorrelation functions of the bead fluctuations in
  equilibrium.  We demonstrate how the results from two different
  experimental runs---one with the protein, and one without---can be
  combined to yield the protein linear response functions.

We apply this two-step procedure on a force clamp setup simulated
through Brownian dynamics, and verify the accuracy of our dynamic
deconvolution method.  Knowledge of mechanical linear response
functions forms the basis of understanding viscoelastic material
properties; the protein case is particularly interesting because every
folding state, i.e. each well in the free energy landscape, will have
its own spectrum of equilibrium fluctuations, and hence a distinct
linear response function.  Two key properties determine this function:
the shape of the free energy around the minimum, and the local
diffusivity.  The latter has contributions both from solvent drag and
the effective roughness of the energy landscape---internal friction
due to molecular torsional transitions and the formation and rupture
of bonds between different parts of the peptide chain.  The
diffusivity profile is crucial for getting a comprehensive picture of
protein folding kinetics and arguably it is just as important as the
free energy landscape itself for very fast folding
proteins~\cite{Cellmer08,Best10,Hinczewski10}.  Our dynamic
deconvolution theory provides a promising route to extract this important
protein characteristic from future force clamp studies.
 
\section{Results and discussion}

\subsection{Force clamp experiments and static deconvolution} 

\begin{figure}[!t]
\centerline{\includegraphics[width=\columnwidth]{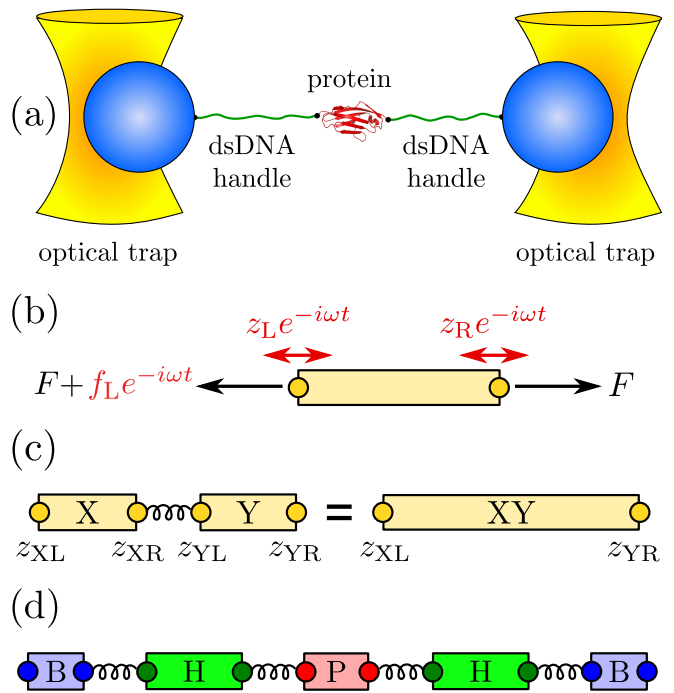}}
\caption{(a) Double optical tweezer force clamp setup for the study of
  equilibrium protein dynamics, with soft traps approximating a
  constant tension $F$. 
   (b) To define linear response functions, consider an individual
  component at tension $F$. A small additional
  oscillatory force $f_\text{L}\exp(-i\omega t)$ applied at the left
  end leads to endpoint oscillations with amplitudes
  $z_L = J_\text{self,L} (\omega) f_\text{L }$ and 
  $z_R = J_\text{cross}(\omega) f_\text{L}$
  which defines the self and cross response functions.
   (c) Two objects X and Y
  connected in series behave as a composite object XY whose response
  functions can be derived through simple rules (Eq.~\eqref{eq1}) from
  the individual X and Y response functions.
  (d) Schematic representation of the optical tweezer setup consisting of 
 beads (B), double-stranded DNA handles (H) and protein
  (P), with connecting springs.  
   }\label{sys}
\end{figure}

As a representative case, in this paper we consider the double trap
setup shown in Fig.~\ref{sys}(a), which typically involves two
optically-trapped polystyrene beads of radius $\sim{\cal
  O}(10^2\:\text{nm})$, two double-stranded DNA handles, each
$\sim{\cal O}(10^2\:\text{nm})$, attached to a protein in the
center\cite{Gebhardt10}.  For fixed trap positions and sufficiently
soft trapping potentials, the entire system will be in equilibrium at
an approximately constant tension $F$.  We are interested in a force
regime ($F \gtrsim 10$ pN in the system under consideration) where the
handles are significantly stretched in the direction parallel to the
applied force (chosen as the $z$ axis), and rotational fluctuations of
the handle-bead contact points are small.  Since the experimental
setup is designed to measure the $z$ separation of the beads as a
function of time, we focus entirely on the dynamic response of the
system along the $z$ direction.  However, the methods below can be
easily generalized to the transverse response as well.  Though we
consider only a passive measurement system in our analysis, an active
feedback loop that minimizes force fluctuations can also be
incorporated, as an additional component with its own characteristic
dynamic response (with the added complication that the response of
the feedback mechanism would have to be independently determined).

To set the stage for our dynamic deconvolution theory, we first
illustrate the static deconvolution for two objects X and Y connected
in series under constant tension $F$, e.g. a protein and a handle.
Let ${\cal P}^\text{X}(z)$ and ${\cal P}^\text{Y}(z)$ be the
constant-force probability distributions for each of these objects
having end-to-end distance $z$.  The total system end-to-end
distribution is given by ${\cal P}^\text{XY}(z) = \int dz^\prime {\cal
  P}^\text{X}(z^\prime) {\cal P}^\text{Y}(z-z^\prime)$.  In terms of
the Fourier-transformed distributions, this can be stated simply
through the ordinary convolution theorem, $\tilde{\cal P}^\text{XY}(k)
= \tilde{\cal P}^\text{X}(k) \tilde{{\cal P}}^\text{Y}(k)$.  If ${\cal
  P}^\text{XY}$ is derived from histograms of the experimental time
series, and if ${\cal P}^\text{Y}$ can be estimated independently
(either from an experiment without the protein, or through theory),
then we can invert the convolution relation to solve for the protein
distribution ${\cal P}^\text{X}$ and thus extract the folding free
energy landscape.  A similar approach works for multiple objects in
series or in parallel.

\subsection{Dynamic response functions}

Before we consider dynamic networks, we define the linear response of
a single object under constant stretching force $F$ along the $z$
direction, as shown in Fig.~\ref{sys}(b).  Imagine applying an
additional small oscillatory force $f_\text{L} \exp(-i\omega t)$ along
the $z$ axis to the left end.  The result will be small oscillations
$z_\text{L} \exp(-i \omega t)$ and $z_\text{R} \exp(-i \omega t)$ of
the two ends around their equilibrium positions.  The complex
amplitudes $z_\text{L}$ and $ z_\text{R}$ are related to $f_\text{L}$
through linear response: $z_\text{L} = J_\text{self,L}(\omega)
f_\text{L}$, $z_\text{R} = J_\text{cross}(\omega) f_\text{L}$,
defining the {\em self response function} $J_\text{self,L}(\omega)$ of
the left end and the {\em cross response function}
$J_\text{cross}(\omega)$.  If the oscillatory force is applied instead
at the right end, the response takes the form: $z_\text{R} =
J_\text{self,R}(\omega) f_\text{R}$, $z_\text{L} =
J_\text{cross}(\omega) f_\text{R}$.  Note that since the object is in
general asymmetric, $J_\text{self,L}(\omega)$ and
$J_\text{self,R}(\omega)$ are distinct functions.  However, there is
only a single cross response (in the absence of time-reversal breaking
effects such as magnetic fields~\cite{Landau}).  For the purposes of
dynamic deconvolution of a network, these three response functions
contain the complete dynamical description of a given component and
are all we need.  It is convenient to define the {\em end-to-end
  response function} $J_\text{ee}(\omega)$, with $ z_\text{R} -
z_\text{L} = J_\text{ee}(\omega)f$, where the oscillatory force $f
\exp(-i\omega t)$ is applied simultaneously to both ends of the object
in opposite directions.  This response turns out to be a linear
combination of the other functions: $J_\text{ee}= J_\text{self,L} +
J_\text{self,R} - 2J_\text{cross}$.

As an illustration we take the simplest, non-trivial example: two
spheres with different mobilities $\mu_\text{L}$ and $\mu_\text{R}$
connected by a harmonic spring of stiffness $k$.  In water we are
typically in the low Reynolds number regime and an overdamped
dynamical description is appropriate.  If an oscillating force of
amplitude $f_\text{L}$ is applied to the left sphere, its velocity
will oscillate with the amplitude $-i\omega z_\text{L} = \mu_\text{L}
(f_\text{L} + k[z_\text{R}-z_\text{L}])$, while the velocity amplitude
of the right sphere is given by $-i\omega z_\text{R} = -\mu_\text{R}
k[z_\text{R}-z_\text{L}]$.  Using the above definitions of the
response functions we obtain
\begin{equation}\label{eq0}
\begin{split}
J_\text{self,L} &= \frac{ \mu_\text{L}(\omega + i \mu_\text{R} k )}{\omega(\mu k - i \omega  )}, 
\:\; J_\text{cross} = \frac{i \mu_\text{L} \mu_\text{R} k}{\omega(\mu k  -i  \omega)},\\
J_\text{ee} &= \frac{\mu}{\mu k -i \omega},
\end{split}
\end{equation}
where $\mu=\mu_\text{L}+\mu_\text{R}$.  By symmetry $J_\text{self,R}$
is the same as $J_\text{self,L}$ with subscripts L and R interchanged.
The end-to-end response $J_\text{ee}$ has a standard Lorentzian form.
For more realistic force transducers, such as semiflexible polymers,
$J_\text{ee}$ will later be written as a sum of Lorentzians reflecting
the polymer normal modes.  Note that when $k=0$, and the spheres no
longer interact, $J_\text{self,L} = i\mu_\text{L}/\omega$, the
standard result for a diffusing sphere, and $J_\text{cross} = 0$, as
expected, since there is no force transmission from one sphere to the
other.

Though all the linear response functions are defined in terms
  of an external oscillatory force, in practice one does not need to
  actually apply such a force to determine the functions
  experimentally.  As described in the two-step deconvolution
  procedure below, one can extract them from measurements that are far
  easier to implement in the lab, namely by calculating
  autocorrelation functions of equilibrium fluctuations.

\subsection{Dynamic convolution of networks} 

Based on the notion of self and cross response functions,
we now consider the dynamics of composites.
We explicitly display the convolution formulas 
for combining two objects in series and in parallel;
by iteration the response of a network of arbitrary topology and complexity can thus be constructed.  As shown in Fig.~\ref{sys}(c),
assume we have two objects X and Y connected by a spring. X is
described by response functions $J^\text{X}_\text{self,L}$,
$J^\text{X}_\text{self,R}$, and $J^\text{X}_\text{cross}$, and we have
the analogous set for Y.  
The internal spring is added for easy evaluation of the force acting between 
the objects, it is eliminated at the end by sending its stiffness to infinity.
We would like to know the response functions of the composite XY object,
$J^\text{XY}_\text{self,X}$, $J^\text{XY}_\text{self,Y}$, and
$J^\text{XY}_\text{cross}$, where the X and Y labels 
correspond to left and right ends, respectively.  The rules
(with full derivation in the Supplementary Information (SI))
 read
\begin{equation}\label{eq1}
\begin{split}
J^\text{XY}_\text{self,X} &= J^\text{X}_\text{self,L} -
 \frac{\left(J^\text{X}_\text{cross}\right)^2}{J^\text{X}_\text{self,R} +
  J^\text{Y}_\text{self,L}},\\
J^\text{XY}_\text{self,Y} &= 
  J^\text{Y}_\text{self,R} - \frac{\left(J^\text{Y}_\text{cross}\right)^2}{J^\text{X}_\text{self,R} + J^\text{Y}_\text{self,L}},\\
 J^\text{XY}_\text{cross} &= \frac{J^\text{X}_\text{cross} J^\text{Y}_\text{cross}}{J^\text{X}_\text{self,R} + J^\text{Y}_\text{self,L}}.
\end{split}
\end{equation}
The rules for connecting two objects in parallel are more
straightforward and read $G^\text{XY}_\alpha = G^\text{X}_\alpha +
G^\text{Y}_\alpha$, where $\alpha$ is any one of the function
categories (self, cross or end-to-end), and $G$ denote the inverse
response functions.  One particularly relevant realization for
parallel mechanical pathways are long-range hydrodynamic coupling
effects, that experimentally act between beads and polymer handles in
the force clamp setup.  We derive the parallel rule and show an
example hydrodynamic application in the SI.  For simplicity, however,
we will concentrate in our analysis on serial connections.  To
proceed, if we set X=H and Y=B, we can obtain the response functions
of the composite handle-bead (HB) object, $J^\text{HB}_\alpha$, if we
know the response functions of the bead and handle separately.  Our
full system in Fig.~\ref{sys}(d) is just the protein sandwiched
between two HB components (oriented such that the handle ends of each
HB are attached to the protein).  The total system response functions
(denoted by ``2HB+P'') in terms of the individual protein and HB
functions result by iterating the pair convolution in Eq.~\eqref{eq1}
twice.  In particular, the end-to-end response
$J^\text{2HB+P}_\text{ee}$ is given by:
\begin{equation}\label{eq1b}
J^\text{2HB+P}_\text{ee} = 2J^\text{HB}_\text{self,B} - \frac{2(J^\text{HB}_\text{cross})^2}{J^\text{HB}_\text{self,H}+J^\text{P}_\text{ee}/2}.
\end{equation}
This is a key relation, since we show below that both
$J^\text{2HB+P}_\text{ee}$ and the three HB response functions
$J^\text{HB}_\text{self,H}$, $J^\text{HB}_\text{self,B}$,
$J^\text{HB}_\text{cross}$ can be derived from force clamp
experimental data.  Hence Eq.~\eqref{eq1b} allows us to estimate the
unknown protein response function $J^\text{P}_\text{ee}$.

We note a striking similarity to the signal processing
scenario~\cite{Oppenheim}, where the output of a linear time-invariant
(LTI) network (e.g. an RLC electric circuit) is characterized through a
``transfer function''.  For such networks, combination rules in terms
of serial, parallel, and feedback loop motifs exist.  The result for
$J^\text{XY}_\text{self,X}$ in Eq.~\eqref{eq1} can be seen in a
similar light: the first term is the self-response
$J^\text{X}_\text{self,X}$ of object X which is independent of the
presence of object Y.  The rational function in the second term is the
``feedback'' due to X interacting with Y.  As expected, if the
cross-response $J^\text{X}_\text{cross}$ connecting the two ends of X
is turned off, this feedback disappears.  In analogy to the transfer
function theory for LTI systems, our convolution rules form a
comprehensive basis to describe the response of an arbitrarily
complicated network inside a force clamp experiment.  And like the
transfer functions which arise out of LTI feedback loops, the
convolution of interacting components consists of a nonlinear
combination of the individual response functions.  The rational
functions due to the feedback of mechanical perturbations across the
connected elements are non-trivial, but can be exactly accounted for
via iteration of the convolution rules.

\subsection{Two-step dynamic deconvolution yields protein dynamic properties}
To illustrate our theory, we construct a two-step procedure to analyze
the experimental system in Fig.~\ref{sys}(a), with the ultimate goal
of determining dynamic protein properties in the force clamp.  Under a
constant force $F$, the protein extension will fluctuate around a mean
corresponding to a folded or unfolded state.  Though it has been
demonstrated that under appropriately tuned forces the protein can
show spontaneous transitions between folded and unfolded states, we
for the moment neglect this more complex scenario.  (In the SI we
analyze simulation results for a protein exhibiting a double-well free
energy, where the two states can be analyzed independently by pooling
data from every visit to a given well; the same idea can be readily
extended to analyze time series data from proteins with one or more
intermediate states.)  We consider the protein dynamics as diffusion
of the reaction coordinate $z^\text{P}_\text{ee}$ (the protein
end-to-end distance) in a free energy landscape
$U_\text{P}(z^\text{P}_\text{ee})$.  The end-to-end response function
$J^\text{P}_\text{ee}$ reflects the shape of $U_\text{P}$ around the
local minimum, and the internal protein friction (i.e the local
mobility), which is the key quantity of interest.  The simplest
example is a parabolic well at position $z^\text{P}_\text{ee} = z_0$,
namely $U_\text{P}(z^\text{P}_\text{ee}) = U_\text{P}(z_0) +
(1/2)k_\text{P} (z^\text{P}_\text{ee}-z_0)^2$.  If we assume the
protein mobility $\mu_\text{P}$ is approximately constant within this
state, the end-to-end response is given by
\begin{equation}\label{eq4}
J^\text{P}_\text{ee}(\omega) = \frac{\mu_\text{P}}{\mu_\text{P} k_\text{P} - i \omega}
\end{equation}
which has the same Lorentzian form as the harmonic two-sphere
end-to-end response in Eq.~\eqref{eq0}.  
Depending on the resolution
and quality of the experimental data, more complex fitting forms may
be substituted, including anharmonic corrections and non-constant
diffusivity profiles (an example of these is given in the
two-state protein case analyzed in the SI).  However for
practical purposes Eq.~\eqref{eq4} is a good starting point.

An experimentalist seeking to determine $J^\text{P}_\text{ee}(\omega)$
would carry out the following two-step procedure:

\begin{figure*}
\begin{center}
\centerline{\includegraphics[width=\textwidth]{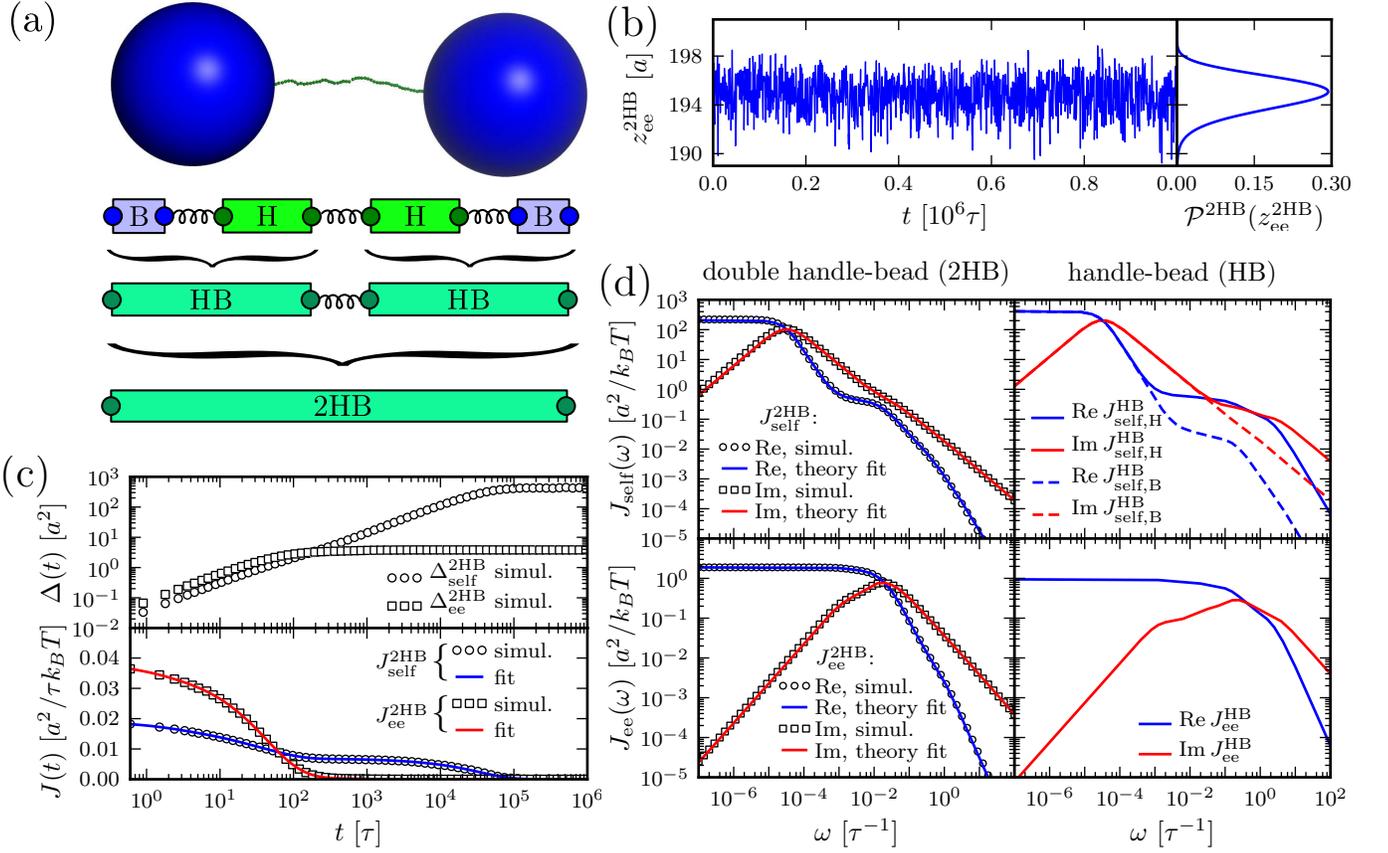}}
\caption{The first step in the deconvolution procedure, needed to to
  determine the handle-bead response $J^\text{HB}$: analysis of the optical tweezer
  system without the protein.  (a) From top to bottom: a Brownian
  dynamics simulation snapshot; schematic representation of the system; 
  after the first convolution step 
  handles and beads are grouped into composite handle-bead (HB) objects;
  after the second convolution step  the full system (``2HB'') constitutes a single object. 
  (b) Part of the
  simulation time series for the total system end-to-end distance
  $z^\text{2HB}_\text{ee}$.  The time series yields the equilbrium
  probability distribution ${\cal P}^\text{2HB}(z^\text{2HB}_\text{ee})$ shown on
  the right.  (c) Top: the MSD functions
  $\Delta^\text{2HB}_\text{self}(t)$ and
  $\Delta^\text{2HB}_\text{ee}(t)$ (Eq.~\eqref{eq5}) calculated from
  the simulation; bottom: the time-domain response functions
  $J^\text{2HB}_\text{self}(t) = (\beta/2)d\Delta^\text{2HB}_\text{self}(t)/dt$,
  $J^\text{2HB}_\text{ee}(t) = (\beta/2)d\Delta^\text{2HB}_\text{ee}(t)/dt$.
  Simulation results (symbols) are numerical derivatives of the curves
  in the top panel.  The solid lines are a 5-exponential fit to the
  simulation results.  (d) Left column: the real and imaginary parts
  of the $J^\text{2HB}$ self and end-to-end response functions.  Simulation
  results (symbols) are just the Fourier transforms of the
  multi-exponential fits in (c).  Theoretical fitting results 
  according to Eq.~\eqref{eq1} and based on HB functions $J^\text{HB}$
  are  shown as solid lines.  Right column: the HB response functions, as
  determined by the theoretical fitting to the full system
  data.}\label{dhb}
\end{center}
\end{figure*}

{\it First step:} Make a preliminary run using a system without the
protein (just two beads and two handles, as illustrated in
Fig.~\ref{dhb}(a)).  As described in the Materials and Methods (MM),
time derivatives of autocorrelation functions calculated from the bead
position time series can be Fourier transformed to directly give
$J^\text{2HB}_\text{self}$ and $J^\text{2HB}_\text{cross}$.  The
convolution rules in Eq.~\eqref{eq1} relate $J^\text{2HB}_\text{self}$
and $J^\text{2HB}_\text{cross}$ to the bead/handle response functions,
$J^\text{HB}_\text{self}$ and $J^\text{HB}_\text{cross}$, which via
another application of Eq.~\eqref{eq1} are related to the response
functions of a single bead and a single handle.  The bead functions
$J^\text{B}_\text{self}$ and $J^\text{B}_\text{cross}$ depend solely
on known experimental parameters (MM), leaving only the handle
functions $J^\text{H}_\text{self}$ and $J^\text{H}_\text{cross}$ as
unknowns in the convolution equations.  Choosing an appropriate
fitting form, determined by polymer dynamical theory (see MM), we can
straightforwardly determine $J^\text{H}_\text{self}$ and
$J^\text{H}_\text{cross}$.

{\it Second step:} Make a production run with the protein.
Eq.~\eqref{eq1b} relates the resulting end-to-end response
$J^\text{2HB+P}_\text{ee}$, extracted from the experimental data, to
the response of the protein alone $J^\text{P}_\text{ee}$.  Since the
first step yielded the composite handle-bead functions
$J^\text{HB}_\text{self,H}$, $J^\text{HB}_\text{self,B}$,
$J^\text{HB}_\text{cross}$ which appear in Eq.~\eqref{eq1b}, the only
unknown is $J^\text{P}_\text{ee}$.  We can thus solve for the
parameters $\mu_\text{P}$ and $k_\text{P}$ which appear in
Eq.~\eqref{eq4}.

This two-step procedure can be repeated at different applied tensions,
revealing how the protein properties (i.e. the intramolecular
interactions that contribute to the diffusivity $\mu_\text{P}$) depend
on force.  Even analyzing the unfolded state of the protein might
yield interesting results: certain forces might be strong enough to
destroy the tertiary structure, but not completely destabilize the
secondary structure, which could transiently refold and affect
$\mu_\text{P}$.

\subsection{Simulations validate the deconvolution technique}
To demonstrate the two-step deconvolution procedure in a realistic
context, we perform Brownian dynamics simulations mimicking a typical force
clamp experiment:  two beads that undergo rotational and translational 
fluctuations  are trapped in 3D harmonic potentials and connected to two 
semiflexible polymers which are linked together via a potential function
that represents the protein folding landscape (see MM for details).
We ignore hydrodynamic effects which can easily be accounted for
 through parallel coupling  pathways, as mentioned above.

We begin with the {\bf first step} of the deconvolution procedure.  A
snapshot of the simulation system, two handles and two beads without a
protein, is shown in Fig.~\ref{dhb}(a). A representative segment of
the $z^\text{2HB}_\text{ee}(t)$ time series is shown in
Fig.~\ref{dhb}(b).  Equilibrium analysis of the time series yields the
end-to-end distribution ${\cal P}^\text{2HB}(z^\text{2HB}_\text{ee})$,
which is useful for extracting static properties of the protein like
the free energy landscape: when the protein is added to the system,
the total end-to-end distribution is just a convolution of the 2HB and
protein distributions.  (The asymmetry of ${\cal
  P}^\text{2HB}$ seen in Fig.~\ref{dhb}(b)
arises from the semiflexible nature of the handles.)  As described in
MM, we use the time series to calculate self and end-to-end MSD curves
$\Delta_\text{self}^\text{2HB}(t)$ and
$\Delta_\text{ee}^\text{2HB}(t)$ [Fig.~\ref{dhb}(c)] whose derivatives
are proportional to the time-domain response functions
$J^\text{2HB}_\text{self}(t)$ and $J^\text{2HB}_\text{ee}(t)$.  The
multi-exponential fits to these functions are illustrated in
Fig.~\ref{dhb}(c), and their analytic Fourier transforms plotted in
the left column of Fig.~\ref{dhb}(d).  We thus have a complete
dynamical picture of the 2HB system response. However, in order to use
Eq.~\eqref{eq1b} to extract the protein response, we first have to
determine the handle-bead response functions $J^\text{HB}_\alpha$.
Although a general fit of $J^\text{HB}_\alpha$ is possible, it is
useful to apply the knowledge about the bead parameters and symmetry
properties of the handle response.  The handle parameters (the set
$\{k^\text{H}_m, \mu_m^\text{H}\}$ in MM Eq.~\eqref{eq3}) are the only
unknowns in the three HB response functions:
$J^\text{HB}_\text{self,H}$, $J^\text{HB}_\text{self,B}$, and
$J^\text{HB}_\text{ee}$.  Note that the handle-bead object is clearly
asymmetric, so the self response will be different at the handle (H)
and bead (B) end.  Convolving two HB objects according to
Eq.~\eqref{eq1} and fitting the handle parameters to the simulation
results for $J^\text{2HB}_\text{self}(\omega)$ and
$J^\text{2HB}_\text{ee}(\omega)$ leads to the excellent description
shown as solid lines on the left in Fig.~\ref{dhb}(d).  The handle
parameters derived from this fitting completely describe the HB
response functions $J^\text{HB}_\alpha$, shown in the right column of
Fig.~\ref{dhb}(d).  The HB response curves reflect their individual
components: there is a low frequency peak/plateau in the
imaginary/real part of the HB self response, related to the slow
relaxation of the bead in the trap.  The higher frequency
contributions are due to the handles, and as a result they are more
prominent in the self response of the handle end than of the bead end.
There is a similarly non-trivial structure in the end-to-end response,
due to the complex interactions between the handle normal modes and
the fluctuations of the trapped bead (see SI for more details).

\begin{figure}[!t]
\centerline{\includegraphics[width=\columnwidth]{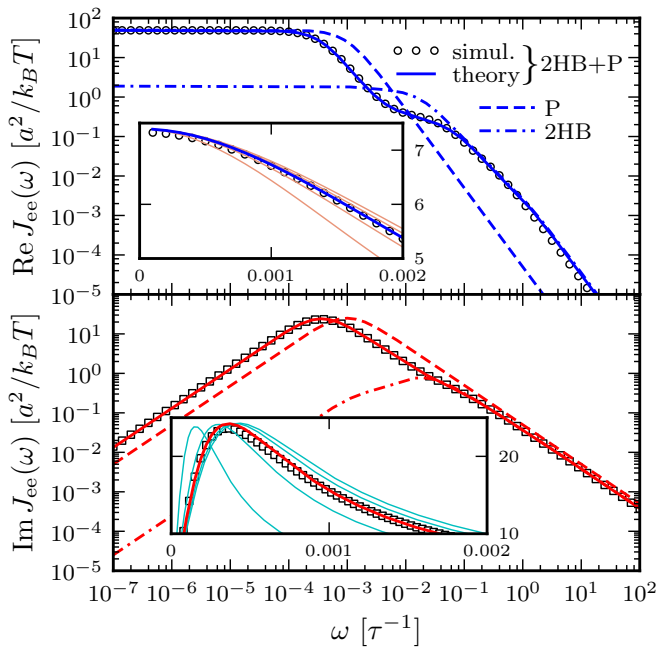}}
\caption{Real (top) and imaginary (bottom) parts of the total
  end-to-end response $J^\text{2HB+P}_\text{ee}(\omega)$ of an optical
  tweezer system with the protein modeled as a single parabolic
  potential well ($k_\text{P} = 0.02$ $k_BT /a^2$, $\mu_\text{P} =
  0.05 \mu_0$).  Symbols are simulation results, and the solid line is
  the theoretical prediction, based on the convolution of the protein
  response $J^\text{P}_\text{ee}(\omega)$ with the HB response
  functions $J^\text{HB}$ of Fig.~\ref{dhb}(d) according to
  Eq.~\eqref{eq1b}.  For comparison, $J^\text{P}_\text{ee}(\omega)$
  (dashed line) and $J^\text{2HB}_\text{ee}(\omega)$ (dot-dashed line)
  are also included.  Insets: to show the sensitivity of the
  theoretical fitting, zoomed-in sections of
  $J^\text{2HB+P}_\text{ee}(\omega)$ near the maxima of the real (top)
  and imaginary (bottom) components. Both simulation (symbols) and
  theoretical (blue/red curve) results are plotted.  The thin
  pink/cyan curves are theoretical results with $\mu_\text{P}$
  different from the true value: from left to right, $\mu_\text{P} =
  0.01$, $0.03$, $0.07$, $0.09 \mu_0$.}\label{sw}
\end{figure}

The double-HB end-to-end distribution in Fig.~\ref{dhb}(b) and the HB
response functions in Fig.~\ref{dhb}(d) are all we need to know about
the optical tweezer system: the equilibrium end-to-end distribution
and linear response of any object which we now put between the handles
can be reconstructed.  We will illustrate this using a toy model of a
protein.  In our simulations for the {\bf second step}, we use a
parabolic potential $U_\text{P}$ with $U^{\prime\prime}_\text{P}(z) =
k_\text{P} = 0.02$ $k_BT/a^2$, and a fixed mobility $\mu_\text{P} =
0.05 \mu_0$.  Here $a=1$ nm and $\mu_0=1/6\pi \eta a$, where $\eta$ is
the viscosity of water.  This leads to the single-Lorentzian response
function $J^\text{P}_\text{ee}$ of Eq.~\eqref{eq4}.  If this exact
theoretical form of $J^\text{P}_\text{ee}$ is convolved with the HB
response functions from the first step according to Eq.~\eqref{eq1b},
we get the result in Fig.~\ref{sw}: very close agreement with
$J^\text{2HB+P}_\text{ee}$ directly derived from the simulated time
series data.  For comparison we also plot the separate end-to-end
responses of the protein alone (P) and the double-HB setup without a
protein (2HB).  As expected $J^\text{2HB+P}_\text{ee}$ differs
substantially from both of these as correctly predicted by the
convolution theory.  The effect of adding handles and beads to the
protein is to shift the peak in the imaginary part of the total system
response to lower frequencies.  Additionally, we see in
$J^\text{2HB+P}_\text{ee}$ the contributions of the handle and bead
rotational motions, which are dominant at higher frequencies.  The
sensitivity of the theoretical fit is shown in the insets of
Fig.~\ref{sw}: zooming in on the maxima of
$\text{Re}\,J^\text{2HB+P}_\text{ee}$ and
$\text{Im}\,J^\text{2HB+P}_\text{ee}$, we plot the true theoretical
prediction (red/blue curves) and results with $\mu_\text{P}$ shifted
away from the the true value (thin pink/cyan curves).  In fact if
$k_\text{P}$ and $\mu_\text{P}$ are taken as free parameters,
numerical fitting to the simulation $J^\text{2HB+P}_\text{ee}$ yields
accurate values of: $\mu_\text{P} = 0.050 \mu_0$ and $k_\text{P} =
0.0199$ $k_BT/a^2$.  Examples of successful deconvolution with other
values of the intrinsic protein parameters are given in the
double-well free energy analysis in the SI.

In practice, any theoretical approach must take into consideration
instrumental limitations: most significantly, there will be a minimum
possible interval $t_\text{s}$ between data collections, related to
the time resolution of the measuring equipment.  The deconvolution
theory can always be applied in the frequency range up to $\omega_s =
1/t_s$.  Whatever physical features of any component in the system
that fall within this range, can be modeled and extracted, without
requiring inaccessible knowledge of fluctuation modes above the
frequency cutoff $\omega_s$.  In the SI, we illustrate this directly on
the toy protein discussed above, coarse-graining the simulation time
series to 0.01 ms intervals, mimicking the equipment resolution used
in Ref.~\cite{Gebhardt10}.  The characteristic frequency of the
protein within the tweezer setup falls within the cutoff, and hence our
two-step deconvolution procedure can still be applied to yield
accurate best-fit results for the protein parameters.  The SI also
includes a discussion of other experimental artifacts---white noise,
drift, and effective averaging of the time series on the time scale
$t_s$---and shows how to adapt the procedure to correct for these
effects.

\section{Conclusion}

Dynamic deconvolution theory allows us to extract the response
functions of a single component from the overall response of a
multicomposite network.  The theory is most transparently formulated
in the frequency domain, and provides the means to reverse the
filtering influence of all elements that are connected to the
component of interest.  From the extracted single-component response
function, dynamic properties such as the internal mobility or friction
can be directly deduced.  At the heart of our theory stands the
observation that the response of any component in the network is
completely determined by three functions, namely the cross response
and the two self responses, which are in general different at the two
ends.  The response of any network can be predicted by repeated
iteration of our convolution formulas for serial and parallel
connections.  Self-similar or more complicated network topologies, as
occur in visco-elastic media, can thus be treated as well.  We
demonstrate the application of our deconvolution theory for a simple
mechanical network that mimics a double-laser-tweezer setup, but the
underlying idea is directly analogous to the signal processing rules
which describe other scalar dynamic networks, such as electrical
circuits or chemical reaction pathways in systems
biology~\cite{Muzzey09}.  We finally point out that dynamic
convolution also occurs in FRET experiments on proteins where
polymeric linkers and conformational fluctuations of fluorophores, as
well as the internal fluorescence dynamics, modify the measured dynamic
fluctuation spectrum~\cite{Gopich09,Chung10,Chung10b}.  The
experimental challenge in the future will thus be to generate
time-series data for single biomolecules with a sufficient frequency
range in order to perform an accurate deconvolution.  For this a
careful matching of the relevant time and spatial scales of the
biomolecule under study and the corresponding scales of the measuring
device (handles as well as beads) is crucial, for which our theory
provides the necessary guidance.

\section{Materials and Methods}

\subsection{Determining the total system response from the experimental time series}  
A key step in the experimental analysis is to obtain the system
response functions $J_\text{self}(\omega)$ and
$J_\text{cross}(\omega)$ from the raw data
(which can either be the  double handle-bead system with or without a protein).
This data consists of two time series $z_\text{B,L}(t)$ and $z_\text{B,R}(t)$
for the left/right bead positions from which we calculate the
mean square displacement (MSD) functions
\begin{equation}\label{eq5}
\begin{split}
\Delta_\text{self}(t) &= \frac{1}{2}\langle (z_\text{B,L}(t) - z_\text{B,L}(0))^2
\rangle \\
&\qquad+ \frac{1}{2}\langle (z_\text{B,R}(t) - z_\text{B,R}(0))^2
\rangle,\\
\Delta_\text{ee}(t) &= \langle (z_\text{ee}(t) - z_\text{ee}(0))^2
\rangle,
\end{split}
\end{equation}
 where $z_\text{ee}(t) = z_\text{B,R}(t) - z_\text{B,L}(t)$, and we
 have averaged the self MSD of the two endpoints because they are
 identical by symmetry.  Calculating the MSD functions is equivalent
 to finding the autocorrelation of the time series: for example, if
 $R_\text{ee}(t) = \langle z_\text{ee}(t) z_\text{ee}(0)\rangle$ is
 the end-to-end autocorrelation, the MSD $\Delta_\text{ee}(t)$ is
 simply given by $\Delta_\text{ee}(t) =
 2(R_\text{ee}(0)-R_\text{ee}(t))$.\\[0.5em] From the
 fluctuation-dissipation theorem~\cite{Landau}, the time-domain
 response functions $J_\text{self}(t)$ and $J_\text{ee}(t)$ are
 related to the derivatives of the MSD functions: $J_\text{self}(t) =
 (\beta/2)d\Delta_\text{self}(t)/dt$, $J_\text{ee}(t) =
 (\beta/2)d\Delta_\text{ee}(t)/dt$, where $\beta = 1/k_B T$. To get
 the Fourier-space response, the time-domain functions can be
 numerically fit to a multi-exponential form, for example
 $J_\text{self}(t) = \sum_i C_i \exp(-\Lambda_i t)$.  In our
 simulation examples typically 4-5 exponentials are needed for a
 reasonable fit.  Once the parameters $C_i$ and $\Lambda_i$ are
 determined, the expression can be exactly Fourier-transformed to give
 the frequency-domain response function, $J_\text{self}(\omega) =
 \sum_i C_i/(\Lambda_i - i\omega)$.  An analogous procedure is used to
 obtain $J_\text{ee}(\omega)$.  The cross response follows as
 $J_\text{cross} = J_\text{self} -J_\text{ee}/2$.  The power spectrum
 associated with a particular type of fluctuation, for example the
 end-to-end spectrum $R_\text{ee}(\omega)$ (defined as the the Fourier
 transform of the autocorrelation), is just proportional to the
 imaginary part of the corresponding response function:
 $R_\text{ee}(\omega) = (2k_B T /\omega)
 \text{Im}\,J_\text{ee}(\omega)$.

\subsection{Bead response functions} 
The response functions of the beads in the optical traps are the
easiest to characterize, since they depend on quantities which are all
known by the experimentalist: the trap stiffness $k_\text{trap}$, bead
radius $R$, mobility $\mu_\text{B} = 1/ 6\pi \eta R$, and rotational
mobility $\mu_\text{rot} = 1 / 8 \pi \eta R^3$.  
Here $\eta$ is the viscosity of water.  For
each bead the three response functions can be defined as described
above, with the two ``endpoints'' being the handle-attachment point on
the bead surface ($z_S$) and the  bead center ($z_B$).  The
latter point is significant because this position is what is directly
measured by the experiment.  For the case of large $F$, where the
rotational diffusion of the bead is confined to small angles away from
the $z$ axis, the response functions are:
\begin{equation}\label{eq2}
\begin{split}
J^\text{B}_\text{self,B}(\omega) &= J^\text{B}_\text{cross}(\omega) = \frac{\mu_\text{B}}{\mu_\text{B}  k_\text{trap} - i \omega},\\
J^\text{B}_\text{self,S}(\omega) &=\frac{\mu_\text{B}}{\mu_\text{B}  k_\text{trap} - i \omega} + 
\frac{2k_B T \mu_\text{rot} R ( F)^{-1}}{2\mu_\text{rot} R  F - i \omega}.
\end{split}
\end{equation}
The second term in $J^\text{B}_\text{self,S}(\omega)$ describes the
contribution of the bead rotational motion, which has a characteristic
relaxation frequency $2\mu_\text{rot}RF$.  This term is derived in the
SI, though it can also be found from an earlier theory of rotational
Brownian diffusion in uniaxial liquid crystals~\cite{Szabo80}.

\subsection{Handle response functions}
The double-stranded DNA handles are semiflexible
polymers whose fluctuation behavior in equilibrium can be decomposed
into normal modes.  We do not need the precise details of this
decomposition, beyond the fact that by symmetry these modes can be
grouped into even and odd functions of the polymer contour length, and
that they are related to the linear response of the polymer through
the fluctuation-dissipation theorem.  From these assumptions, the
handle response functions have the following generic form (a fuller
description can be found in the SI):
\begin{equation}\label{eq3}
\begin{split}
J^\text{H}_\text{self}(\omega) &= \frac{i \mu^\text{H}_0}{\omega} + 
\sum_{n=1}^{N_\text{mode}} \frac{\mu_n^\text{H} }{\mu_n^\text{H} k^\text{H}_n - i \omega},\\
J^\text{H}_\text{cross}(\omega) &= \frac{i \mu^\text{H}_0}{\omega} + 
\sum_{n=1}^{N_\text{mode}} \frac{(-1)^m \mu_n^\text{H} }{\mu_n^\text{H} k^\text{H}_n  - i \omega},
\end{split}
\end{equation}
for some set of $2N_\text{mode}+1$ parameters
$\{\mu_n^\text{H},k^\text{H}_n\}$.  Note that since the handles are
symmetric objects, the self response of each endpoint is the same
function $J^\text{H}_\text{self}(\omega)$.  The mobilities
$\mu^\text{H}_n$ and elastic coefficients $k^\text{H}_n$ encode the
normal mode characteristics, with the mode relaxation times $\tau_n
\equiv 1/(\mu_n^\text{H} k^\text{H}_n)$ ordered from largest ($n=1$)
to smallest ($n=N_\text{mode}$). The parameter $\mu_0^\text{H}$ is the
center-of-mass mobility of the handle along the force direction.
Simple scaling expressions for the zeroth and first mode parameters in
terms of physical polymer parameters as well as the connection between
the expressions in Eq.~\eqref{eq0} and Eq.~\eqref{eq3} are given in
the SI.  (These same expressions, with a smaller $l_p$, could describe
a completely unfolded, non-interacting, polypeptide chain at high
force.) In practice, the high-frequency cutoff $N_\text{mode}$ can be
kept quite small (i.e. $N_\text{mode}=4$) to describe the system over
the frequency range of interest.

\subsection{Numerical inversion of convolution equations}
Care must be taken in manipulating Fourier-space relationships like
Eq.~\eqref{eq1b}.  Directly inverting such equations generally leads
to numerical instabilities due to noise and singularities.  In our
case, we can avoid direct inversion because the forms of the component
functions are known beforehand (i.e. Eq.~\eqref{eq2} for the beads,
Eq.~\eqref{eq3} for the handles, Eq.~\eqref{eq4} for the protein).
Thus when we model the response of the double handle-bead system with
a protein, $J^\text{2HB+P}_\text{ee}(\omega)$, we end up through
Eqs.~\eqref{eq1} and \eqref{eq1b} with some theoretical function
$\tilde{J}^\text{2HB+P}_\text{ee}(\omega,\mb{K})$ where $\mb{K}$ is
the set of unknown parameters related to the components.  Since
$J^\text{2HB+P}_\text{ee}(\omega)$ is known as a function of $\omega$
from the experimental time series, we find $\mb{K}$ by minimizing the
goodness-of-fit function $M(\mb{K}) = \sum_{\omega \in \Omega}
\left[\log|J^\text{2HB+P}_\text{ee}(\omega)| - \log|
  \tilde{J}^\text{2HB+P}_\text{ee}(\omega,\mb{K}) |\right]^2$, where
$\Omega$ is a logarithmically spaced set of frequencies, up to the
cutoff frequency $\omega_\text{s} = 1/t_\text{s}$ determined by the
time resolution $t_\text{s}$ of the measuring equipment. This is
equivalent to simultaneously fitting the real and imaginary parts of
our system response on a log-log scale.

\vspace{1em}
\subsection{Simulations}
In our Brownian dynamics simulations each handle
is a semiflexible bead-spring chain of 25 beads of radius $a = 1$ nm,
every bead having mobility $\mu_0 = 1/6\pi \eta a$.  The handle
persistence length is $l_p = 20a$.  The harmonic springs used to
connect all components together (including the beads making up the
handles) have stiffness $\gamma = 300$ $k_B T/a^2$.  The
beads  have radius $R = 50a$, and the traps have
strength $k_\text{trap} = 0.00243$ $k_B T/a^2$, which corresponds to
0.01 pN/nm.  The traps are positioned such that the average force in
equilibrium $F \approx 3\:k_BT/a = 12.35$ pN.\\[0.5em]
To capture the essential features of protein dynamics, we construct a
simple toy model.  The protein is characterized by two vectors: a
center-of-mass position $\mb{r}^\text{P}_\text{cm}$ and an end-to-end
separation $\mb{r}^\text{P}_\text{ee}$.  Both
$\mb{r}^\text{P}_\text{cm}$ and the transverse components of
$\mb{r}^\text{P}_\text{ee}$ obey standard Langevin dynamics with a
mobility $\mu_\text{cm} = \mu _{\perp} = 0.12 \mu_0$.  The internal
dynamics of protein fluctuations is modeled through the longitudinal
end-to-end component $z^\text{P}_\text{ee}$, subject to a potential
$U_\text{P}(z^\text{P}_\text{ee})$  and a mobility
$\mu_\text{P}$.  The transverse components
$(x^\text{P}_\text{ee}, y^\text{P}_\text{ee})$ feel a harmonic
potential with spring constant $k_\perp=1.5$ $k_BT/a^2$.\\[0.5em]
The simulation dynamics are implemented through a discretized
free-draining Langevin equation with time step $\Delta t = 3\times
10^{-4}\,\tau$, where the time unit $\tau = a^2/k_B T \mu_0$.  Data is
collected every 1000 time steps.  Typical simulation times are $\sim
{\cal O}(10^{10})$ steps, with averages collected from $\approx 20$
independent runs for each system considered.

\begin{acknowledgments}
This work was supported by Deutsche Forschungsgemeinschaft (DFG)
within grant SFB 863.  We thank the Feza G\"ursey Institute for
computational resources provided by the Gilgamesh cluster.
\end{acknowledgments}

\newpage

\begin{widetext}

\renewcommand{\theequation}{S.\arabic{equation}}
\renewcommand{\thefigure}{S.\arabic{figure}}
\setcounter{equation}{0}
\setcounter{figure}{0}
\setcounter{section}{0}

{\centering \bf \large Supplementary Material for ``Deconvolution of dynamic mechanical networks'' \\}

\section{Derivation of dynamic convolution equations}\label{series}

Consider a system under tension $F$ consisting of two objects X and Y
connected by a spring of stiffness $\gamma$, as shown in Fig.~1(c) in
the main text.  The left and right equilibrium endpoint positions are
$Z_\text{XL}$, $Z_\text{XR}$ respectively for X, and $Z_\text{YL}$,
$Z_\text{YR}$ for Y.  Imagine applying a small additional oscillatory
force $f_\text{L} \exp(-i \omega t)$ at $Z_\text{XL}$ along the $z$
direction.  In the long time limit, every endpoint coordinate in the
system would exhibit oscillations of the form $z_{\alpha i}
\exp(-i \omega t)$ with some amplitude $z_{\alpha i}$, $\alpha =
\text{X},\text{Y}$, $i=L,R$.  Let us also denote the instantaneous
force exerted by the connecting spring as $f \exp(-i \omega t)$.  In
terms of $f$ and the four amplitudes $z_{\alpha i}$, we can
write down five equations based on the definitions of the self/cross
response functions for X and Y:
\begin{equation}\label{eq:s1}
\begin{split}
f &= \gamma( z_\text{YL}- z_\text{XR}),\\
 z_\text{XL} &= J^\text{X}_\text{self,L} f_\text{L} + J^\text{X}_\text{cross} f,\\
 z_\text{XR} &= J^\text{X}_\text{cross} f_\text{L} + J^\text{X}_\text{self,R} f,\\
 z_\text{YL} &= -J^\text{Y}_\text{self,L} f,\\
 z_\text{YR} &= -J^\text{Y}_\text{cross}f,
\end{split}
\end{equation}
where the $\omega$ dependence of the response functions is implicit.
We would like to relate the X and Y response functions to those of
the full system.  Since the external force perturbation is applied at
the X end of the system, by definition $ z_\text{XL} =
J^\text{XY}_\text{self,X} f_\text{L}$, $ z_\text{YR} =
J^\text{XY}_\text{cross} f_\text{L}$.  Solving for $
z_\text{XL}$ and $ z_\text{YR}$ from Eq.~\eqref{eq:s1}, we find
the following expressions for the full system response functions in
the limit $\gamma \to \infty$:
\begin{equation}\label{eq:s2}
J^\text{XY}_\text{self,X} = J^\text{X}_\text{self,L} - \frac{\left(J^\text{X}_\text{cross}\right)^2}{J^\text{X}_\text{self,R} + J^\text{Y}_\text{self,L}},\qquad J^\text{XY}_\text{cross} = \frac{J^\text{X}_\text{cross} J^\text{Y}_\text{cross}}{J^\text{X}_\text{self,R} + J^\text{Y}_\text{self,L}}.
\end{equation}
If the force perturbation is applied at YR instead of XL, an analogous
derivation yields the remaining self response function
$J^\text{XY}_\text{self,Y}$:
\begin{equation}\label{eq:s3}
J^\text{XY}_\text{self,Y} = J^\text{Y}_\text{self,R} - \frac{\left(J^\text{Y}_\text{cross}\right)^2}{J^\text{X}_\text{self,R} + J^\text{Y}_\text{self,L}}.
\end{equation}
By iterating Eqs.~\eqref{eq:s2}-\eqref{eq:s3}, reducing each pair of
components into a single composite object, we can readily obtain the
convolution equations for an arbitrary number of components in series.

\section{Dynamic convolution with parallel pathways and long-range hydrodynamic interactions}

Though the systems described in the main text consist of multiple
objects connected in series, one can extend the theory to cases where
parallel connections are also present.  The most significant
experimental realization of such connections are objects coupled
through long-range hydrodynamic interactions.  To make our convolution
theory comprehensive, in this section we will derive the rules for
parallel pathways, and then show in particular how they can be used to
treat hydrodynamics.

\subsection{Dynamic convolution for two objects in parallel}\label{parallel}

Following the notation of SI Sec.~\ref{series}, consider two objects X
and Y in parallel (in other words, sharing the same left and right
end-points).  For object $\alpha = \text{X},\text{Y}$, let $f_{\alpha\text{L}}
\exp(-i \omega t)$ and $f_{\alpha\text{R}} \exp(-i \omega t)$ be the
forces that need to be applied on that object at the left and right
ends in order that the end-points exhibit oscillations $ z_{\alpha \text{L}}
\exp(-i \omega t)$ and $ z_{\alpha \text{R}}\exp(-i \omega t)$.  The
relationship between the end-point force and oscillation amplitudes is
given by:
\begin{equation}\label{par1}
\begin{split}
 z_{\alpha\text{L}} &= J^\alpha_\text{self,L} f_{\alpha\text{L}} + J^\alpha_\text{cross} f_{\alpha\text{R}},\\
 z_{\alpha \text{R}} &= J^\alpha_\text{cross} f_{\alpha \text{L}} + J^\alpha_\text{self,R} f_{\alpha\text{R}}.
\end{split}
\end{equation}
Since the objects are connected in parallel, we know that $
z_\text{XL} = z_\text{YL} \equiv z_\text{L}$ and $ z_\text{XR} =
z_\text{YR} \equiv z_\text{R}$.  Hence we can write Eq.~\eqref{par1}
as a matrix equation,
\begin{equation}\label{par1b}
\mb{z} = \overleftrightarrow{J}^\alpha \mb{f}_\alpha,
\end{equation}
 where:
\begin{equation}\label{par2}
\mb{z} = \begin{pmatrix}  z_\text{L} \\  z_\text{R} \end{pmatrix}, \quad
\mb{f}_\alpha = \begin{pmatrix} f_{\alpha\text{L}} \\ f_{\alpha\text{R}} \end{pmatrix}, \quad
\overleftrightarrow{J}^\alpha = \begin{pmatrix} J^\alpha_\text{self,L} & J^\alpha_\text{cross} \\ J^\alpha_\text{cross} &  J^\alpha_\text{self,R} \end{pmatrix}.
\end{equation}
Looking at the full system of two objects together, the total forces
on the left and right ends required for a particular set of
oscillation amplitudes $ z_\text{L}$ and $ z_\text{R}$ are additive: $
\mb{f} = \mb{f}_\text{X} + \mb{f}_\text{Y}$, where $\mb{f} =
(f_\text{L},f_\text{R})$.  If we define the inverse response matrix,
$\overleftrightarrow{G}^\alpha \equiv
(\overleftrightarrow{J}^\alpha)^{-1}$, then
\begin{equation}\label{par2b}
\mb{f}_\alpha = \overleftrightarrow{G}^\alpha \mb{z}
\end{equation}
and we obtain:
\begin{equation}\label{par2c}
\mb{f} = (\overleftrightarrow{G}^\text{X} + \overleftrightarrow{G}^\text{Y}) \mb{z} \equiv \overleftrightarrow{G}^\text{XY} \mb{z}.
\end{equation}
This is the main
result for convolving parallel pathways: the inverse response matrices
of each pathway additively combine to yield the total inverse response
matrix. Let us label the components of the inverse response matrix
$G^\alpha$ as follows:
\begin{equation}\label{par3}
\overleftrightarrow{G}^\alpha = (\overleftrightarrow{J}^\alpha)^{-1} =\begin{pmatrix} \frac{J^\alpha_\text{self,R}}{J^\alpha_\text{self,L}J^\alpha_\text{self,R} - (J^\alpha_\text{cross})^2} &
  -\frac{J^\alpha_\text{cross}}{J^\alpha_\text{self,L}J^\alpha_\text{self,R} - (J^\alpha_\text{cross})^2} \\ -\frac{J^\alpha_\text{cross}}{J^\alpha_\text{self,L}J^\alpha_\text{self,R} - (J^\alpha_\text{cross})^2} &
  \frac{J^\alpha_\text{self,L}}{J^\alpha_\text{self,L}J^\alpha_\text{self,R} - (J^\alpha_\text{cross})^2} \end{pmatrix} \equiv \begin{pmatrix} G^\alpha_\text{self,L} &
  G^\alpha_\text{cross} \\ G^\alpha_\text{cross} &
  G^\alpha_\text{self,R} \end{pmatrix}.
\end{equation}
Then since $\overleftrightarrow{J}^\text{XY} = (\overleftrightarrow{G}^\text{XY})^{-1}$, we can solve for the XY
response functions in terms of the components of $\overleftrightarrow{G}^\text{XY} =
\overleftrightarrow{G}^\text{X} + \overleftrightarrow{G}^\text{Y}$:
\begin{equation}\label{par4}
\begin{split}
&J^\text{XY}_\text{self,L} = \frac{G^\text{XY}_\text{self,R}}{G^\text{XY}_\text{self,L}G^\text{XY}_\text{self,R} - (G^\text{XY}_\text{cross})^2},
\quad J^\text{XY}_\text{cross} = -\frac{G^\text{XY}_\text{cross}}{G^\text{XY}_\text{self,L}G^\text{XY}_\text{self,R} - (G^\text{XY}_\text{cross})^2},\\
&\qquad\qquad\qquad\qquad J^\text{XY}_\text{self,R} = \frac{G^\text{XY}_\text{self,L}}{G^\text{XY}_\text{self,L}G^\text{XY}_\text{self,R} - (G^\text{XY}_\text{cross})^2}.
\end{split}
\end{equation}
Eqs.~\eqref{par3}-\eqref{par4} constitute the complete rules for dynamic convolution of two objects in parallel.

\subsection{Incorporating hydrodynamics as a parallel connection}

As a simple, realistic application of the above parallel convolution
rules, consider the following setup, which constitutes two parallel
pathways: a) two beads of self-mobility $\mu_\text{B}$ trapped in
optical tweezers of strength $k_\text{trap}$, interacting through
long-range hydrodynamics; b) a DNA handle which connects the two
beads.  (The handle can actually be any composite object in series,
for example two DNA strands on either side of a protein, like in the
main text.)  Though in principle all the objects in the system are
coupled hydrodynamically to each other, we will consider only a single
pairwise interaction: between the two beads.  Since the beads are the
objects with the largest drag, this is by far the most important
interaction for the practical purposes of analyzing experimental data.
Though the rotational degrees of freedom for the beads can be
incorporated in the theory, we will ignore them in this example, since
they give only a small correction at large forces.  Thus the
center-of-mass of bead $i$, and the handle end-point attached to that
bead, will have the same oscillation amplitude: $ z_i$,
$i=\text{L},\text{R}$.  

To get the total system behavior, we start by looking at each pathway
independently, and calculating the associated inverse response matrix.
If we assume the handle is symmetric with end-point response functions
$J^\text{H}_\text{self}$ and $J^\text{H}_\text{cross}$, the matrix
$\overleftrightarrow{G}^\text{H}$ for the handle pathway follows from Eq.~\eqref{par3}:
\begin{equation}\label{par5}
\overleftrightarrow{G}^\text{H} =\begin{pmatrix} \frac{J^\text{H}_\text{self}}{(J^\text{H}_\text{self})^2 - (J^\text{H}_\text{cross})^2} & -\frac{J^\text{H}_\text{cross}}{(J^\text{H}_\text{self})^2 - (J^\text{H}_\text{cross})^2} \\ -\frac{J^\text{H}_\text{cross}}{(J^\text{H}_\text{self})^2 - (J^\text{H}_\text{cross})^2} & \frac{J^\text{H}_\text{self}}{(J^\text{H}_\text{self})^2 - (J^\text{H}_\text{cross})^2} \end{pmatrix} \equiv \begin{pmatrix} G^\text{H}_\text{self} &
  G^\text{H}_\text{cross} \\ G^\text{H}_\text{cross} &
  G^\text{H}_\text{self} \end{pmatrix}.
\end{equation}
For the bead pathway, we can derive $\overleftrightarrow{G}^\text{B}$ starting from the
equations of motion of the two trapped beads in the presence of
external force amplitudes $f_\text{BL}$ and $f_\text{BR}$ acting on
the left and right beads:
\begin{equation}\label{par6}
\begin{split}
-i\omega  z_\text{L} &= \mu_\text{B,self} f^\text{tot}_\text{BL}
+ \mu_\text{B,cross} f^\text{tot}_\text{BR},\\ 
-i\omega  z_\text{R} &= \mu_\text{B,self} f^\text{tot}_\text{BR} + \mu_\text{B,cross} f^\text{tot}_\text{BL}.
\end{split}
\end{equation}
Here $f^\text{tot}_{\text{B}i}$ is the {\it total} force amplitude on
the $i$th bead, including the contribution of the trap force and the
external force: $f^\text{tot}_{\text{B}i} = -k_\text{trap}  z_i
+ f_{\text{B}i}$.  The terms $\mu_\text{B,self}$ and
$\mu_\text{B,cross}$ are respectively the diagonal and off-diagonal
components of the two-bead hydrodynamic mobility tensor.
$\mu_\text{B,self}$ is the bead self-mobility, and
$\mu_\text{B,cross}$ is the cross mobility, describing the long-range
hydrodynamic coupling between the beads.  We assume both of these
mobilities are roughly constant.  (The justification for this
approximation is given at the end of this section, along with full
expressions for $\mu_\text{B,self}$ and $\mu_\text{B,cross}$ in a
typical experimental setup.)  We can solve Eq.~\eqref{par6} for
$f_{\text{B}i}$, expressing it in the form $\mb{f}_\text{B} =
\overleftrightarrow{G}^\text{B} \mb{z}$, where:
\begin{equation}\label{par7}
\overleftrightarrow{G}^\text{B} = \begin{pmatrix} k_\text{trap} - \frac{i \omega \mu_\text{B,self}}{(\mu_\text{B,self})^2 - (\mu_\text{B,cross})^2}  & \frac{i\omega \mu_\text{B,cross}}{(\mu_\text{B,self})^2 - (\mu_\text{B,cross})^2} \\ \frac{i\omega \mu_\text{B,cross}}{(\mu_\text{B,self})^2 - (\mu_\text{B,cross})^2} & k_\text{trap} - \frac{i \omega \mu_\text{B,self}}{(\mu_\text{B,self})^2 - (\mu_\text{B,cross})^2} \end{pmatrix} \equiv \begin{pmatrix} G^\text{B}_\text{self} &
  G^\text{B}_\text{cross} \\ G^\text{B}_\text{cross} &
  G^\text{B}_\text{self} \end{pmatrix}.
\end{equation}
The full system inverse response matrix is $\overleftrightarrow{G}^\text{HB} = \overleftrightarrow{G}^\text{H} + \overleftrightarrow{G}^\text{B}$.  Starting with Eq.~\eqref{par5} for $\overleftrightarrow{G}^\text{H}$ and
Eq.~\eqref{par7} for $\overleftrightarrow{G}^\text{B}$, we can derive the components of the
total response $\overleftrightarrow{J}^\text{HB}$ from Eq.~\eqref{par4}.

To complete the description of this system, we need expressions for
$\mu_\text{B,self}$ and $\mu_\text{B,cross}$.  In equilibrium, the
beads have some average center-to-center separation $\Delta Z$, and
for added realism, we will consider the bead centers to be a height
$H$ above a surface (i.e. the microscope slide).  In a typical setup
$\Delta Z$ and $H$ may be on the order of the bead radius $R$,
resulting in non-negligible modifications to each bead's self-mobility
from both the presence of the wall (assumed to be a no-slip boundary),
and the presence of the other bead.  Though the mobilities will
instantaneously depend on the exact positions of the beads relative to
themselves and the wall, we will use the mobility values at the
equilibrium positions, since the amplitude of the bead fluctuations in
the traps is small compared to $\Delta Z$ and $H$.  Hence the
mobilities will be functions of $\Delta Z$, $H$, and $R$.  The bead
self mobility can be estimated as:
\begin{equation}\label{par8}
\mu_\text{B,self} = \mu_\text{B} \left(1 - \frac{9R}{16H} + \frac{R^3}{8H^3} \right)\left( 1 - \frac{15 R^4}{4 {\Delta Z}^4} \right),
\end{equation}
where $\mu_\text{B} = 1/(6 \pi \eta R)$ is the self-mobility of a bead
alone in an unbounded fluid.  The first parenthesis is the correction
due to the wall, and the second parenthesis is the
correction due to the other bead~\cite{KimKarrila}.  Finally, for
$\mu_\text{B,cross}$ we use the corresponding component of the Blake
tensor~\cite{Blake71} (at the Rotne-Prager level), which describes the
cross mobility of two beads moving parallel to a no-slip boundary:
\begin{equation}\label{par9}
\begin{split}
\mu_\text{B,cross} &= \frac{\mu_\text{B} R}{2} \left[ 2R^2 \left( \frac{16H^4 - 22H^2 {\Delta Z}^2 + {\Delta Z}^4}{(4H^2 + {\Delta Z}^2)^{7/2}} - \frac{1}{{\Delta Z}^3}\right)\right.\\
&\qquad +\left. 3 \left(\frac{1}{{\Delta Z}}-\frac{12H^4+4H^2{\Delta Z}^2+{\Delta Z}^4}{(4H^2+{\Delta Z}^2)^{5/2}}\right)\right].
\end{split}
\end{equation}

\section{Details of the bead response functions}

In the absence of bead rotation (where the bead only has
center-of-mass translational degrees of freedom), the self response
functions of both the bead center ($z_\text{C}$) and and
handle-attachment point on the bead surface ($z_\text{S}$) are
identical: they describe a sphere with translational mobility
$\mu_\text{B}$ in an optical trap of strength $k_\text{trap}$, namely
$J_\text{self,B}^\text{B}(\omega) = J_\text{self,S}^\text{B}(\omega) =
\mu_\text{B}/(\mu_\text{B} k_\text{trap}-i\omega)$.  (Here, as in the
main text, we ignore self-mobility corrections due to the microscope
slide surface or the presence of the other bead; if desired, these can
be simply incorporated as described in the previous section.)
Similarly, since the bead is a rigid body and any force perturbation
at the bead center is directly communicated to the bead surface,
$J^\text{B}_\text{cross}(\omega)$ is equal to
$J_\text{self,B}^\text{B}(\omega)$.

The situation gets more complicated when the rotational degrees of
freedom are included, but this involves only the surface response
function $J^\text{B}_\text{self,S}$, which now has to include an extra
term to account for the rotation of $z_\text{S}$, along with the
translational motion of the bead itself.  To derive the rotational
response (i.e. the second term of $J^\text{B}_\text{self,S}$ in
Eq.~[6] of the main text) we take advantage of the
fluctuation-dissipation theorem.  If $\mb{R}$ is the vector connecting
the bead center to the handle-attachment point, such that $R_z = z_S -
z_B$, and $\mb{u}(t) = \mb{R}(t)/R$ the corresponding unit vector,
then we would like to calculate the time correlation function $C_R(t)
= \langle R_z(t) R_z(0)\rangle - \langle R_z\rangle^2 = R^2 (\langle
u_z(t)u_z(0)\rangle - \langle u_z \rangle^2)$ in the case where $z_S$
is subject to a constant force $F$ along the $z$ axis.  Here $\langle
R_z\rangle$ refers to the equilibrium average of $R_z$.  For the
purposes of analyzing the optical tweezer experiment, we consider only
the large force case, where $F \gg k_B T/R$, and thus $u_z(t)$ is in
the vicinity of 1.  The fluctuation-dissipation theorem states that
the time-domain rotational response function is equal to $-\beta
dC_R(t)/dt$, where $\beta = 1/k_B T$, which we can Fourier transform
to get the frequency-domain response.

The key to calculating $C_R(t)$ is to note that the equilibrium
fluctuations of $\mb{u}(t)$ correspond to diffusion on the surface of
a unit sphere subject to the external potential $V(\mb{u}) = -FRu_z$.
Let us define the Green's function $G(\mb{u},\mb{u}^\prime;t)$ as the
conditional probability of ending up at $\mb{u}$ at time $t$, given an
initial position $\mb{u}^\prime$ at $t=0$, or equivalently
$G(\mb{u},\mb{u}^\prime;0) = \delta(\mb{u}-\mb{u}^\prime)$.  This
Green's function satisfies the Smoluchowski equation for rotational
diffusion~\cite{DoiEdwards}:
\begin{equation}\label{bd1}
\frac{\partial}{\partial t}G(\mb{u},\mb{u}^\prime;t) = \mu_\text{rot}{\cal R}\cdot \left({\cal R} G(\mb{u},\mb{u}^\prime;t) + k_BT G(\mb{u},\mb{u}^\prime;t) {\cal R}V(\mb{u})\right),
\end{equation}
where ${\cal R} \equiv \mb{u} \times \partial/\partial\mb{u}$.  The
correlation $\langle u_z(t) u_z(0) \rangle$ can be expressed in terms
of $G$ as:
\begin{equation}\label{bd2}
\langle u_z(t) u_z(0) \rangle = \int d^2\mb{u} \int d^2\mb{u}^\prime \, u_z u_z^\prime G(\mb{u},\mb{u}^\prime;t) \Psi_\text{eq}(\mb{u}^\prime),
\end{equation}
where $\Psi_\text{eq}(\mb{u}) = A \exp(-\beta V(\mb{u}))$ is the
equilibrium probability distribution of $\mb{u}$, with normalization
constant $A$.  Taking the time derivative of both sides of
Eq.~\eqref{bd2}, and substituting the right-hand side of
Eq.~\eqref{bd1} for $\partial G/\partial t$, one can derive the
following relation describing the time evolution of
$C_R(t)$~\cite{Coffey}:
\begin{equation}\label{bd3}
\frac{d}{d t}C_R(t) = -2k_BT\mu_\text{rot} C_R(t) - \frac{2}{3}\mu_\text{rot}F D_R(t).
\end{equation}
Here $D_R(t)$ is the higher order correlation function $D_R(t) = R^3
(\langle P_2(u_z(t)) u_z(0) \rangle - \langle P_2(u_z)\rangle \langle
u_z \rangle)$, where $P_2(x) = (3x^2-1)/2$ is the second-order
Legendre polynomial.  In fact, $d D_R(t)/d t$ can itself be expressed
in terms of even higher order correlation functions, and this
procedure can be iterated to form an infinite set of linear
differential equations.  Rather than formally solving this set, which
is possible but tedious, we employ the effective relaxation time
approximation~\cite{Coffey}, where $C_R(t)$ is assumed to be dominated
by a single exponential decay with relaxation time $\tau$, namely
$C_R(t) \approx C_R(0) \exp(-t/\tau)$.  The coefficient $C_R(0) =
R^2(\langle u_z^2\rangle - \langle u_z\rangle^2)$, and can be
evaluated using the equilibrium distribution $\Psi_\text{eq}$,
yielding $C_R(0) = (k_BT)^2/F^2$ for large $F$.  Using
Eq.~\eqref{bd3}, the relaxation time can be written as:
\begin{equation}\label{bd4}
\tau^{-1} = -\frac{1}{C_R(0)} \frac{d}{dt}C_R(0) = \frac{1}{C_R(0)}\left[2k_BT\mu_\text{rot}C_R(0) + \frac{2}{3}\mu_\text{rot}F D_R(0)\right].
\end{equation}
Like $C_R(0)$, the $D_R(0)$ appearing on the right of Eq.~\eqref{bd4}
is an equilibrium average determined through $\Psi_\text{eq}$: $D_R(0)
= 3(k_B T)^2 R/F^2$ in the large $F$ limit.  Since in this limit the
first term in the brackets in Eq.~\eqref{bd4} is negligible compared
to the second, we ultimately find $\tau^{-1} = 2\mu_\text{rot}RF$.
(The same approximate expression for the relaxation time $\tau$ can
also be derived from a theory describing rotational Brownian diffusion
in uniaxial liquid crystals~\cite{Szabo80}.)  The rotational response
is the Fourier transform of $-\beta dC_R/dt = \beta C_R(0) \tau^{-1}
\exp(-t/\tau)$:
\begin{equation}\label{bd5}
\frac{\beta C_R(0) \tau^{-1}}{\tau^{-1} -i\omega} = \frac{2k_B T
  \mu_\text{rot}R/F}{2\mu_\text{rot}RF-i\omega}.
\end{equation}

\section{Details of the handle response functions}

We can use standard ideas from the theory of polymer
dynamics~\cite{DoiEdwards} to derive a simple fitting form for the
handle response functions.  Consider a handle in isolation at
equilibrium under constant tension $F$, described by a continuous
space curve with $z$ component $z(s,t)$ at time $t$ and contour
coordinate $s$, where $s$ runs from 0 to $L$.  Without deriving a
detailed dynamical theory, one can still make a few generic
assumptions about the behavior of $z(s,t)$.  The first is that it can
be decomposed into a sum over normal modes $\Psi_n(s)$ in the
following way:
\begin{equation}\label{eqh1}
z(s,t) = z_0(s,t) + \sum_{n=1}^\infty P_n(t) \Psi_n(s).
\end{equation}
Here the first term $z_0(s,t)$ is some reference contour, and the
second term represents deviations from that reference contour, where
$P_n(t)$ is the coefficient of the $n$th normal mode.  For convenience
we set the reference contour $z_0(s,t) = \langle z(s,t) \rangle$, the
thermodynamic average over all polymer configurations in equilibrium.
The $t$ dependence of $z_0(s,t)$ incorporates the center-of-mass
motion of the polymer, so that $\langle (z_0(s,t)-z_0(s,0))^2 \rangle
= 2D_\text{cm} t$, where $D_\text{cm}$ is the center-of-mass diffusion
constant.  With this choice of reference contour, $\langle P_n(t)
\rangle =0$.

The second assumption is that the equilibrium time correlations of the
normal mode coefficients have the form,
\begin{equation}\label{eqh2}
\langle P_n(t) P_m(0) \rangle = \delta_{n,m} A_n \exp(-\lambda_n t),
\end{equation}
for $t \ge 0$, where $A_n$ is some constant, and $\lambda_n$ is the
inverse relaxation time of the $n$th normal mode.  This type of
exponentially decaying normal mode correlation appears in dynamical
theories for many types of polymers: flexible
chains~\cite{DoiEdwards}, nearly rigid rods~\cite{Granek1997},
mean-field approximations for semiflexible
chains~\cite{Harnau1996,Hinczewski2009}.  By convention, we order the
normal modes such that $\lambda_n$ increases with $n$, i.e. $n=1$
corresponds to the largest relaxation time.  The third and final
assumption is that due to the symmetry of the handle (the two
end-points are equivalent), the normal modes $\Psi_n(s)$ can be
grouped into even and odd functions of $s$ around the center $L/2$,
such that $\Psi_n(L) = (-1)^n \Psi_n(0)$.

Putting all these properties together, we can derive expressions for
the MSD of a handle end-point, $\Delta^\text{H}_\text{self}(t) =
\langle (z(L,t)-z(L,0))^2\rangle$, and the cross MSD,
$\Delta^\text{H}_\text{cross}(t) = \langle (z(L,t)-z(0,0))^2\rangle$:
\begin{equation}\label{eqh3}
\begin{split}
\Delta^\text{H}_\text{self}(t) &= 2 D_\text{cm} t + 2\sum_{n=1}^{\infty} A_n (1-e^{-\lambda_n t}) \Psi^2_n(L),\\
\Delta^\text{H}_\text{cross}(t) &= 2 D_\text{cm} t + 2\sum_{n=1}^{\infty} (-1)^n A_n (1-e^{-\lambda_n t}) \Psi^2_n(L).
\end{split}
\end{equation}
From the fluctuation-dissipation theorem, the time-domain response
functions are related to the MSDs through: $J^\text{H}_\alpha(t) =
(\beta/2)d\Delta_\alpha(t)/dt$, $\alpha = \text{self}$, $\text{cross}$.
Taking the Fourier transforms of $J^\text{H}_\alpha(t)$ yields the
handle fitting forms shown in Eq.~[7] of the main text:
\begin{equation}\label{eqh4}
\begin{split}
J^\text{H}_\text{self}(\omega) &= \frac{i \mu^\text{H}_0}{\omega} + \sum_{n=1}^{\infty} \frac{\mu^\text{H}_n}{\mu^\text{H}_n k^\text{H}_n - i \omega},\\
J^\text{H}_\text{cross}(\omega) &= \frac{i \mu^\text{H}_0}{\omega} + \sum_{n=1}^{\infty} \frac{(-1)^n \mu^\text{H}_n}{\mu^\text{H}_nk^\text{H}_n - i \omega},
\end{split}
\end{equation}
where $\mu^\text{H}_0 = \beta D_\text{cm}$, $\mu^\text{H}_n = \beta
A_n \lambda_n \Psi^2_n(L)$, $k_n^\text{H} = \lambda_n/\mu^\text{H}_n$,
$n>0$.  In calculating the Fourier transforms, we note that all terms
in Eq.~\eqref{eqh3} are implicitly multiplied by the unit step
function $\Theta(t)$, since the MSD functions are defined only for
$t\ge 0$.

As a demonstration of the handle fitting forms, consider the simplest
case, where a handle consists of two spheres, like in the response
function example preceding Eq.~[1] in the main text.  The functions in
Eq.~[1] can be rewritten in terms of a center-of-mass and one normal
mode contribution, as expected for a single-spring system:
\begin{gather}\label{eqh5}
J_\text{self,L} = \frac{i \mu_\text{L} \mu_\text{R}/\mu}{\omega} + \frac{\mu_\text{L}^2/\mu}{\mu k - i \omega}, \quad J_\text{self,R} = \frac{i \mu_\text{L} \mu_\text{R}/\mu}{\omega} + \frac{\mu_\text{R}^2/\mu}{\mu k - i \omega},\nonumber\\
J_\text{cross} = \frac{i \mu_\text{L} \mu_\text{R}/\mu}{\omega} - \frac{\mu_\text{L}\mu_\text{R}/\mu}{\mu k - i \omega}. 
\end{gather}
In the symmetric case where $\mu_\text{L} = \mu_\text{R} = \mu/2$,
these response functions have exactly the same form as Eq.~\ref{eqh4},
with $\mu^\text{H}_0 = \mu^\text{H}_1 = \mu/4$, $k^\text{H}_1 = 4k$.

For polymer handles used in experiments there will be contributions
from a spectrum of normal modes.  However the response functions are
still dominated by the center-of-mass and lowest-frequency normal mode
terms.  Hence for the purpose of estimating the linear response
characteristics of a given setup, we provide simple scaling
expressions for the parameters $\mu^\text{H}_0$, $\mu^\text{H}_1$, and
$k^\text{H}_1$ in the case of a handle which is a semiflexible polymer
of contour length $L$ and persistence length $l_p$
(i.e. double-stranded DNA).  

By analogy to the two-sphere example above, $\mu^\text{H}_0 =
\mu^\text{H}_1$ is just the center-of-mass mobility of the handle
along the stretching direction, and $k^\text{H}_1 = 4k$, where $k$ is
the effective spring constant of the handle.  For large forces, where
the handle is near maximum extension, we can approximate it as thin
rod, for which the mobility $\mu_\parallel$ along the long axis is
given by $\mu_\parallel = \ln(L/d)/(2\pi \eta L)$ to leading order.
Here $\eta$ is the viscosity of water and $d=2a$ the diameter of the
handle.  Hence $\mu^\text{H}_0 = \mu^\text{H}_1 \approx
\mu_\parallel$.  To get the effective spring constant, the starting
point is the approximate Marko-Siggia interpolation
formula~\cite{Marko95} relating the tension $F$ felt by the handle to
its average end-to-end extension $z_\text{ee}$ along the $z$ axis:
\begin{equation}\label{eqh6}
F = \frac{k_BT}{l_p}\left[\frac{z_\text{ee}}{L} + \frac{1}{4(1-z_\text{ee}/L)^2} - \frac{1}{4}\right].
\end{equation}
Since the force magnitude $F$ is related to the polymer free energy
${\cal F}$ through $F = \partial {\cal F}/\partial z_\text{ee}$, and
the effective handle spring constant $k = \partial^2 {\cal F}/\partial
z_\text{ee}^2 = \partial F/\partial z_\text{ee}$, we can estimate $k$
from Eq.~\eqref{eqh6}:
\begin{equation}\label{eqh7}
\begin{split}
k = \frac{k_BT}{l_p}\left[\frac{1}{L} + \frac{1}{2L(1-z_\text{ee}/L)^3}\right] &=  \frac{k_BT}{l_pL}\left[1 + 4\left(\frac{l_p F}{k_B T} -\frac{z_\text{ee}}{L} +\frac{1}{4}\right)^{3/2}\right]\\
&\approx \frac{4 k_BT}{l_pL}\left(\frac{l_p F}{k_B T}\right)^{3/2},
\end{split}
\end{equation}
where in the second line we have used the fact that $F \gg k_B T/l_p$
for the cases we consider, i.e. $k_B T/l_p \approx 0.1$ pN for
double-stranded DNA, where $l_p \approx 50$ nm.  Since $k_1^\text{H} =
4k$, we can derive the following expression for the longest relaxation
time $\tau_1 = (\mu_1^\text{H} k_1^\text{H})^{-1}$ of the handle
end-to-end fluctuations along the force direction (in other words the
relaxation time of the first normal mode):
\begin{equation}\label{eqh8}
\tau_1 = \frac{\pi\eta l_p L^2}{8k_BT\ln(L/d)}\left(\frac{l_p F}{k_B T}\right)^{-3/2}.
\end{equation}
Up to a constant prefactor, this is the large force limiting case of
an earlier scaling expression for the longitudinal relaxation time
that has been verified through optical tweezer experiments on single
DNA molecules~\cite{Meiners00}.

With the parameter values $a=1$ nm, $T = 298$ K, and $\eta = 0.89$
mPa$\cdot$s, we can estimate a typical relaxation time for DNA handles
where $l_p = 50$ nm and $L \sim {\cal O}(100\:\text{nm})$: $\tau_1
\sim {\cal O}(10\:\text{ns})$.  For the specific case of the Brownian
dynamics simulations, where the handles are somewhat shorter at
$L=50a$, the scaling argument predicts $\tau_1 \approx 0.7\tau = 3$
ns, comparable to the numerically fitted value $\tau_1
\approx 0.25\tau = 1$ ns, where $\tau = a^2/k_B T \mu_0 = 4$ ns.  In
Fig.~2(d) in the main text, the frequency scale $\tau_1^{-1} \approx
4\tau^{-1} = 10^9$ s$^{-1}$ is where we see the clear divergence
between the handle-end and bead-end HB self response functions
$J^\text{HB}_\text{self,H}(\omega)$ and
$J^\text{HB}_\text{self,B}(\omega)$.  This is related to the fact that
the handle fluctuations become the dominant contribution for the HB
object above this frequency scale.  (Given the bead radius $R = 50a$
and trap strength $k_\text{trap} = 0.00243$ $k_BT/a^2$, the
characteristic bead frequency scale $\mu_B k_\text{trap} \approx
5\times 10^{-5}\tau^{-1} = 12500$ s$^{-1}$ is much lower.)  As a
general design principle, experiments will have the best
signal-to-noise characteristics when there is a good separation of
scales between the handle, protein, and bead characteristic
frequencies.  The example in the main text satisfies this idea, since
the protein characteristic frequency $\mu_\text{P} k_\text{P} =
10^{-3}\tau^{-1} = 2.5\times 10^5$ s$^{-1}$ is distinct from the
ranges of the beads and handles.

\section{Details of the individual handle and bead contributions to the handle-bead response functions}

\begin{figure}[!t]
\centerline{\includegraphics[scale=1]{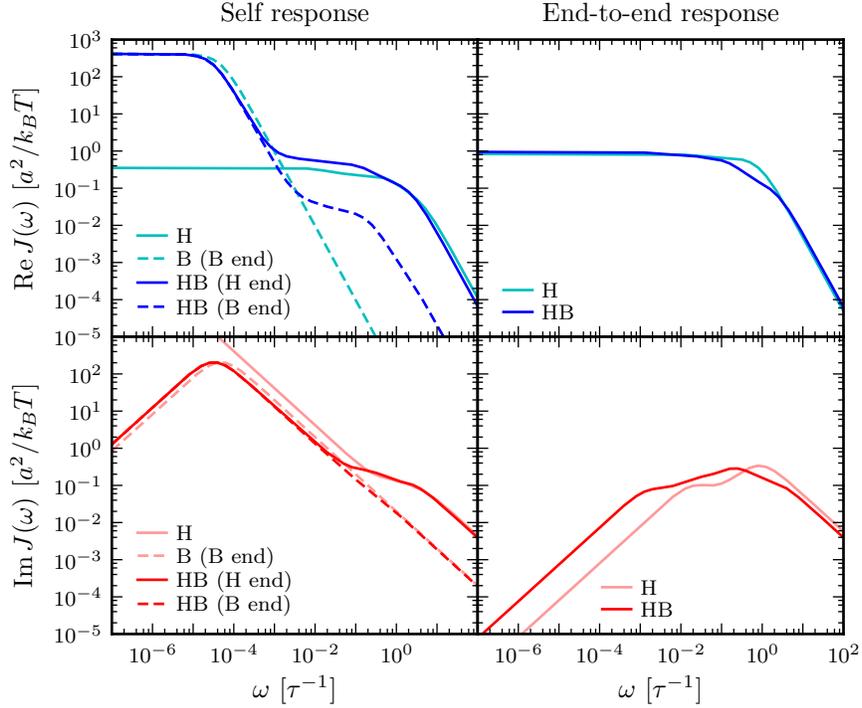}}
\caption{The panels show the handle-bead (HB) response functions,
  taken from Fig. 2 in the main text, along with functions for the
  handle (H) and bead (B) separately.  The self and end-to-end
  response functions are in the left and right columns respectively,
  with real parts in the top row and imaginary parts in the bottom
  row.}\label{hb}
\end{figure}

As described in the main text, the HB response functions reflect the
contributions of their handle and bead components.  To elucidate this,
we plot in Fig.~\ref{hb} the HB functions from Fig.~2, together with
the individual H and B functions.  As expected, the HB self response
at the H end is very similar to the individual H response, and at the
B end it is close to the individual B response.  For the HB end-to-end
function, only comparison with the handle component is possible (since
the bead end-to-end response is zero).  Again the two functions are
similar, but clearly adding the bead onto the handle perturbs its
end-to-end response, shifting it to lower frequencies and changing its
shape.  The way in which the components contribute to the response of
the total object is not merely additive, and requires the convolution
rules in order to be to accurately predicted.

\section{Dynamic deconvolution of a double-well protein}

\begin{figure}[!t]
\centerline{\includegraphics[width=\textwidth]{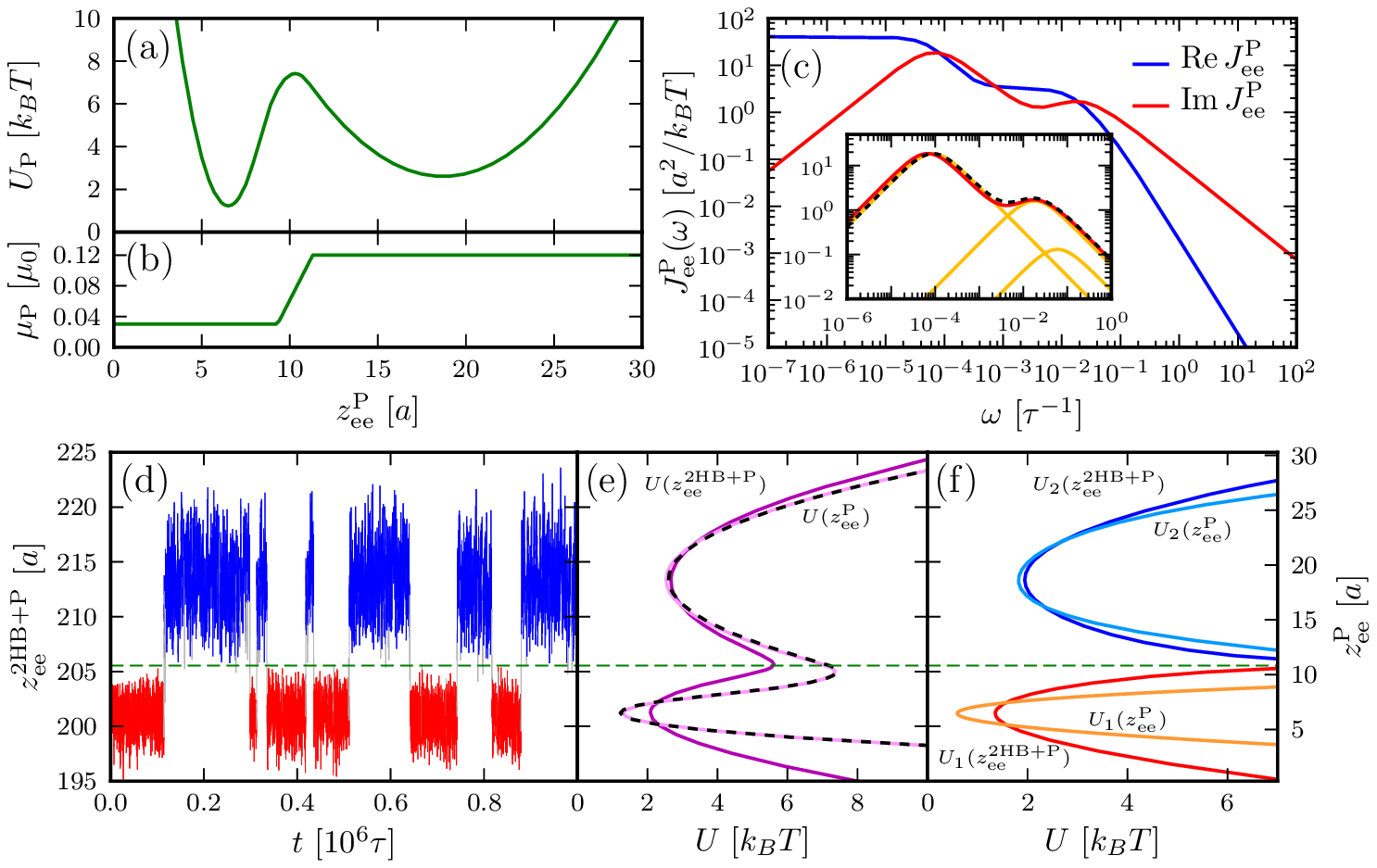}}
\caption{Analysis of an optical tweezer system with the protein
  modeled using a double-well potential.  (a) The protein potential
  $U_\text{P}$ as a function of end-to-end distance
  $z^\text{P}_\text{ee}$, from Eq.~\eqref{eq6}.  The potential as
  shown is tilted by the external stretching force $F = 3$ $k_B T/a$.
  (b) The protein mobility profile
  $\mu_\text{P}(z^\text{P}_\text{ee})$.  (c) Formal linear response
  result for $J^\text{P}_\text{ee}(\omega)$ based on a numerical
  Fokker-Planck solution using the full profiles in (a) and (b).
  Inset: $\text{Im}\,J^\text{P}_\text{ee}(\omega)$ (red curve), with
  analytical estimates for the three contributions to the protein
  response (orange).  The sum of these analytical estimates is shown
  as a dashed black curve.  (d) Fragment of time series for
  $z^\text{2HB+P}_\text{ee}$, with the portions assigned to either the
  folded or unfolded wells colored red and blue respectively.  The
  remainder is shaded gray.  (e) The full system free energy profile
  $U(z^\text{2HB+P}_\text{ee})$ calculated from the simulation (dark
  purple), together with the protein-only result
  $U(z^\text{P}_\text{ee})$ (light purple) obtained by static
  deconvolution using the 2HB distribution in Fig.~2(b) of the main
  text.  This agrees with the original theoretical form,
  Eq.~\eqref{eq6}, shown as a dashed line.  (f) For the time series in
  each well, the corresponding full system potential
  $U_i(z^\text{2HB+P}_\text{ee})$ (blue/red lines) and the deconvolved result
  $U_i(z^\text{P}_\text{ee})$ (cyan/orange lines), $i=1,2$.}\label{dw}
\end{figure}

Instead of the single-well protein of the main text, we start with a
double-well potential $U_\text{P}(z^\text{P}_\text{ee})$ shown in
Fig.~\ref{dw}(a).  This potential arises from the force:
\begin{equation}\label{eq6}
F_\text{P}(z) = \begin{cases} -k_1(z-z_1) & z\le \frac{k_m z_m+k_1 z_1}{k_1+k_m}\\
 k_m(z-z_m) & \frac{k_m z_m+k_1 z_1}{k_1+k_m}< z \le \frac{k_m z_m+k_2 z_2}{k_2+k_m}\\
-k_2(z-z_2) & \frac{k_m z_m+k_2 z_2}{k_2+k_m} < z \end{cases},
\end{equation}
where the parameters $k_1 = 2\:k_BT/a^2$, $k_m = 1.5\:k_BT/a^2$, $k_2
= 0.15\:k_BT/a^2$, $z_1 = 5a$, $z_m= 12.3a$, $z_2=-1.3a$.  The energy
profile plotted in Fig.~\ref{dw}(a) has been additionally tilted
by a contribution $F z$ representing the equilibrium stretching at
constant force $F=3\:k_BT/a$.  To mimic the effects of internal
friction on the protein dynamics, we allow for a coordinate-dependent
mobility $\mu_\text{P}(z^\text{P}_\text{ee})$, which is plotted in
Fig.~\ref{dw}(b).  In the left well $\mu_\text{P} = 0.03\:\mu_0$
(folded state), while in the right well we have a higher value
$\mu_\text{P} = 0.12\:\mu_0$ (unfolded state), with a linear
transition of width $2a$ around the energy barrier.  For real
proteins, larger internal friction in the folded state may occur due
to a greater proportion of intact native bonds compared to unfolded
configurations.

Formally, the protein end-to-end response
$J^\text{P}_\text{ee}(\omega)$ may be derived for this double-well
potential using a numerical solution of the corresponding
Fokker-Planck equation, based on the method of Bicout and
Szabo~\cite{Bicout98}.  The numerical procedure consists of the
following steps.  We discretize our potential $U_\text{P}(z_i)$ over
$M$ values $z_i$, $i=1,\ldots,M$, in some range $z_\text{min}$ to
$z_\text{max}$.  The bin size $d = (z_\text{max}-z_\text{min})/M$.
Define local equilibrium probabilities $p_i =
\exp(-U_\text{P}(z_i)/k_BT)/Z$, where $Z=\sum_i
\exp(-U_\text{P}(z_i)/k_BT)$.  Transition frequencies $w(i|j)$ of
going from bin $i$ to bin $j$ are given by:
\begin{equation}\label{eq:bs1}
w(i|j) = k_B T \frac{\mu_\text{P}(z_i)+\mu_\text{P}(z_j)}{2d^2}
\exp\left(-\frac{\beta(U_\text{P}(z_i)-U_\text{P}(z_j))}{2}\right).
\end{equation}
Reflective boundary conditions at the ends of our range are imposed by
defining $w(0|1) = w(1|0) = w(M|M+1) = w(M+1|M) = 0$.  The $M\times M$
symmetric rate matrix $R$ has the nonzero elements:
\begin{equation}\label{eq:bs2}
R_{i,i} = -w(i+1|i) - w(i-1|i), \qquad R_{i,i\pm 1} = R_{i\pm 1,i} = \sqrt{w(i\pm 1|i)w(i|i\pm 1)}.
\end{equation}
The eigenvalues and eigenvectors of this matrix satisfy an equation of
the form: $R \mb{v}_\alpha = -\Lambda_\alpha \mb{v}_\alpha$, where the
index $\alpha = 0,\ldots,M-1$ and we arrange the eigenvalues in
increasing order.  With reflective boundary conditions at either end,
the smallest eigenvalue is always $\Lambda_0 = 0$.  The correlation
function $C^\text{P}_\text{ee}(t) = \langle z_\text{ee}^\text{P}(t)
z_\text{ee}^\text{P}(0)\rangle$ for the protein is then given by
$C^\text{P}_\text{ee}(t) = \sum_{\alpha=1}^{M-1} C_\alpha
e^{-\Lambda_\alpha t}$, where the coefficients $C_\alpha$ are:
\begin{equation}\label{eq:bs3}
C_\alpha = \sum_{i=1}^{M} \left(\sqrt{p_i} z_i v_{\alpha i} \right)^2.
\end{equation}
Here $v_{\alpha i}$ is the $i$th component of the eigenvector
$\mb{v}_\alpha$.  Finally, from the fluctuation-dissipation theorem,
the time-domain response function $J^\text{P}_\text{ee}(t)$ is related
to $C^\text{P}_\text{ee}(t)$ through $J^\text{P}_\text{ee}(t) =
-\beta dC^\text{P}_\text{ee}(t)/dt$.  After a Fourier transform, we get
the response function:
\begin{equation}\label{eq:bs4}
J^\text{P}_\text{ee}(\omega) = \sum_{\alpha=1}^{M-1} \frac{\beta C_\alpha \Lambda_\alpha}{\Lambda_\alpha - i \omega},
\end{equation}
where all the parameters $\{C_\alpha,\Lambda_\alpha\}$ are known
numerically.

This response function is shown in Fig.~\ref{dw}(c).  Its structure
can actually be decomposed into three contributions, as shown in the
inset: two correspond to fluctuations of the protein about the local
minima; the third, at lower frequencies, corresponds to transitions
between the wells.  The orange curves in the inset are simple
analytical predictions for these contributions: the two intrawell
peaks at higher frequencies are just single Lorentzians with the
appropriate values of $k$ and $\mu_\text{P}$ for each well, and the
interwell transition peak is a based on a discrete two-state
description~\cite{Wio99}.  Though approximate, the analytical sum
(dashed line) captures well the exact numerical result (red line).

However if we go beyond this and try to naively apply the convolution
theory on the $J^\text{P}_\text{ee}(\omega)$ of Fig.~\ref{dw}(c), the
results for the total system will deviate from the actual behavior as
seen in simulations.  This is not surprising, since the (large
amplitude, low frequency) transitions between wells are a
fundamentally nonlinear process, and thus violate the linear response
assumption on which our theory is based.  To correctly apply our
analysis in this situation, we should focus on the linear response of
the system within each state, where the protein fluctuates about the
local minimum: in essence, we break the problem into two single-well
systems.

To implement this approach, we need to extract ``single-well'' time
series from the full simulation data, a portion of which is shown in
Fig.~\ref{dw}(d).  Since what we measure is the end-to-end behavior of
the total system, we choose the separation line between wells to lie
at the barrier position in the full system free energy profile
$U(z^\text{2HB+P}_\text{ee}) = -k_B T \log {\cal
  P}(z^\text{2HB+P}_\text{ee})$ [Fig.~\ref{dw}(e)].  Segments of the
time series falling on either side of the separation line will be
assigned to well 1 (folded) or well 2 (unfolded), with the following
exceptions: points that are within a certain time window $\pm \delta
t$ of a transition (i.e. a crossing of the separation line) are
excluded.  Setting $\delta t$ to $10^6 \tau$, longer than the
relaxation times within the wells, this exclusion ensures that the
resulting single-well time series are not significantly perturbed by
memory effects from the transitions.  Note that times both before and
after each transition are excluded in order to preserve the
time-reversal symmetry of the equilibrium time series.  For the time
series in Fig.~\ref{dw}(d), the parts assigned to well 1 and 2 are
colored red and blue respectively, with the remainder shaded gray.
The same method can be generalized to a more complicated free energy
landscape, for example in a protein with intermediate states, to get a
single-well time series corresponding to every state.

The end-to-end distributions from the single-well series yield
corresponding energy profiles $U_1(z^\text{2HB+P}_\text{ee})$ and
$U_2(z^\text{2HB+P}_\text{ee})$, shown in Fig.~\ref{dw}(f).  After
static deconvolution with the double-HB distribution from
Fig.~2(b), we get profiles in terms of the protein end-to-end
distance: $U_1(z^\text{P}_\text{ee})$ and $U_2(z^\text{P}_\text{ee})$.
Near the minima these are identical to the original $U_\text{P}$, but
have anharmonic corrections as they approach infinity at the
separation line.  The dynamic deconvolution analysis for each well is
analogous to the simple parabolic case described in the main text,
except instead of the single Lorentzian we will use the corrected
$J^\text{P}_\text{ee}$ based on the Fokker-Planck numerical solution
with $U_1(z^\text{P}_\text{ee})$ and $U_2(z^\text{P}_\text{ee})$.
This yields a generalized form in each well:
$J^\text{P}_\text{ee}(\omega) = \sum_{\alpha} \frac{\beta C_\alpha
  \Lambda_\alpha}{\Lambda_\alpha - i \omega}$, with the multiple
Lorentzians accounting for the anharmonic corrections.  For the
parabolic case, $C_1 = k_B T/k$, $\Lambda_1 = \mu_\text{P} k$, and there
were no other terms.  Given a $\mu_\text{P}$ for the well, the
parameters $\{C_\alpha, \Lambda_\alpha\}$ are known from the
Fokker-Planck solution.  (In fact only the $\Lambda_\alpha$ depend on
the diffusion constant; note that $J^\text{P}_\text{ee}(0) =
\sum_\alpha \beta C_\alpha = \beta \langle (z^\text{P}_\text{ee})^2\rangle$ is an
equilibrium average, so the $C_\alpha$ must be independent of
$\mu_\text{P}$).

\begin{figure}[!t]
\centerline{\includegraphics[width=8cm]{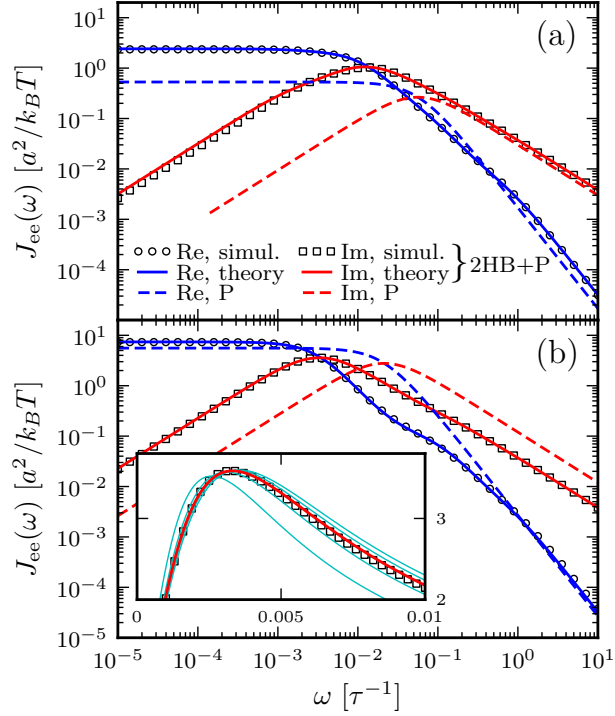}}
\caption{Analogous to Fig.~3 in the main text, showing the accuracy of
  the theoretical prediction of the total system end-to-end response
  $J^\text{2HB+P}_\text{ee}(\omega)$ using the double-well protein
  model.  The (a) folded and (b) unfolded states are analyzed
  separately, with the simulation results plotted as symbols and the
  theory as solid lines.  The latter uses the true values of
  $\mu_\text{P} = 0.03$ $\mu_0$ in the folded state and $\mu_\text{P}
  = 0.12$ $\mu_0$ in the unfolded state.  Blue/red results denote
  the Re/Im parts respectively.  For comparison, the protein response
  $J^\text{P}_\text{ee}(\omega)$ (dashed line) is also shown for
  each well, based on the local Fokker-Planck numerical solution.
  Inset to (b): for the unfolded state, a zoomed-in section of the
  $\text{Im}\,J^\text{2HB+P}_\text{ee}(\omega)$ peak, with simulation
  (symbols) and theoretical (red curve) results.  The cyan curves are
  theoretical results with $\mu_\text{P}$ different from the true value:
  from left to right, $\mu_\text{P} = 0.04$, $0.08$, $0.016$, $0.24$
  $\mu_0$.}\label{dw2}
\end{figure}

As in the single-well example, applying the convolution of
$J^\text{P}_\text{ee}$ with the HB response functions, and comparing
with the simulation results, shows excellent agreement
[Fig.~\ref{dw2}].  Here the theoretical curves (red lines) for each
well are based on setting $\mu_\text{P}$ to the values at the minima.
For the unfolded state, the inset in Fig.~\ref{dw2} shows how this
theoretical result shifts if $\mu_\text{P}$ is displaced from the true
value of $0.12\:\mu_0$.  The sensitivity of the theoretical fitting
allows one to numerically extract good estimates for $\mu_\text{P}$ in
each well: $\mu_\text{P} = 0.033\:\mu_0$ (folded) and $\mu_\text{P} =
0.111\:\mu_0$ (unfolded), where the exact values are $0.03\:\mu_0$
and $0.12\:\mu_0$ respectively. Given the complications associated with
excluding interwell hopping (and the nonconstant diffusion profile
across the barrier), it is remarkable that we can still fit to within
$\approx 10\%$ of the true values.  If one was interested in getting
just an estimate of $\mu_\text{P}$ in each well, without necessarily
getting a perfect fit for the total system response, the anharmonic
corrections could be ignored and the single Lorentzian form used
instead of the exact $J^\text{P}_\text{ee}$ in the fitting.  The
resulting values for $\mu_\text{P}$ have a comparable accuracy.

\section{Accounting for experimental limitations:  noise, drift, and finite time resolution}

Practical implementation of the deconvolution theory on experimental
time series must take into account possible errors due to instrumental
limitations.  We will focus on the optical tweezer apparatus in
particular, since this is the example analyzed in the main text.
However the techniques discussed below are applicable in a generic
experimental setup.  For optical tweezers, effects like drift and
noise which enter into the measured output are typically divided into
``experimental'' and ``Brownian'' categories~\cite{Moffitt08}: the
former arises from the instrumental components, while the latter is
related to thermal fluctuations of the beads and the objects under
study.  Equilibrium Brownian noise is completely captured within the
deconvolution theory: the linear response formalism provides a
quantitative prediction of exactly how each thermally fluctuating part
of the system will contribute to the total signal.  We will thus
concentrate on experimental effects, and outline how they enter into
the theoretical deconvolution procedure.

For a perfect instrument, the raw data collected by the
experimentalist would be a ``true'' time series $z_\alpha(t_i)$ at
discrete time steps $t_i$, where $\alpha$ refers to either one of the
bead positions ($\alpha = \text{B,L}$ or $\text{B,R}$), or the
bead-bead separation ($\alpha = \text{ee}$).  The sampling time $t_\text{s}
\equiv t_{i+1} - t_i$ is the time resolution of the apparatus.  In any
realistic scenario, the actual measured time series is
$\tilde{z}_\alpha(t_i) = z_\alpha(t_i) + \eta_\alpha(t_i)$, where
$\eta_\alpha(t_i)$ is an extraneous signal due to the instrument.  We
will consider two possible contributions to $\eta_\alpha(t)$: a) white
noise, or any high-frequency random signal that is uncorrelated at
time scales above $t_\text{s}$; b) a low-frequency signal related to
instrumental drift over an extended measurement period.  The latter
can result from environmental factors like slow temperature changes
and air currents that affect the laser beam position.  As a side note,
one of the advantages of the dual-trap setup illustrated in the text
is a dramatic reduction of such drift effects, relative to earlier
designs involving a single trap and biomolecules tethered to a
surface~\cite{Moffitt08}.  Whereas the beam position of a single trap
will move over time relative to the surface, the dual traps are
created from a single laser, and hence any overall drift in the beam
position will not affect the trap separation.  (Though smaller issues
may exist, like the effective mobility of the beads shifting as the
distance from the sample chamber surface varies.)  In any case, the
low drift of the dual-trap setup allows for long, reliable data
collection periods (i.e. on the order of a minute in the leucine
zipper study of Ref.~\cite{Gebhardt10}).

The final experimental effect we will describe in this section relates
to the time resolution: the sampling period $t_\text{s}$ cannot be made
smaller than a certain value, dependent on the instrumentation.  Not
only is the frequency of data collection limited, but because the
electronics have a finite response time, there will also effectively be
some kind of averaging of the true signal across the $t_\text{s}$ time window
between measurements.  Both of these effects can be incorporated into
the deconvolution analysis, and we will illustrate this concretely
through the toy model protein simulation data discussed in the main
text.  These simulations had data collection at intervals of $0.3 \tau
= 1.2$ ns, in order to clearly illustrate that the theory is valid
over a wide frequency regime encompassing the characteristic
fluctuations of all the system components.  However we can mimic the
experimental apparatus by resampling and averaging the data over
windows of size $t_\text{s} = 0.01$ ms (the resolution of the Gebhardt {\it
  et. al.}~\cite{Gebhardt10} setup).  The deconvolution procedure
works for the reduced frequency range (up to $\omega_s = t_\text{s}^{-1}$),
and the main quantity of interest (the protein diffusivity) can still
be extracted with good accuracy.  The reason for this is that the
protein characteristic frequency in our example (after being slowed
down by the handles and beads) falls under the cutoff $\omega_s$.  In
general, any component fluctuation modes with frequencies $\omega
\lesssim \omega_s$ can be deconvolved using the theory, even though we
lack information about higher frequency modes above the
cutoff.  Though we may not be able to directly observe the modes of
certain objects---like the short DNA handles whose characteristic
frequencies are $\sim {\cal O}(1\:\text{ns})$---this does not impede us
from exploiting the full physical content of the time series in the
frequency range below $\omega_s$.

We begin our detailed discussion with the white noise and drift
effects that contribute to $\eta_\alpha(t)$.

\subsection{White noise and drift}

Consider the situation where our time series is contaminated by a
Gaussian white noise signal $\eta_\alpha(t_i)$, with zero mean and
correlation $\langle \eta_\alpha(t_i) \eta_\alpha(t_j) \rangle =
\gamma \delta_{ij}$.  We will also assume that the white noise is to
a good approximation not correlated with the true component signal:
$\langle \eta_\alpha(t_i) z_\alpha(t_j) \rangle = 0$, $\forall i,j$.
From the measured time series $\tilde{z}_\alpha(t_i) = z_\alpha(t_i) +
\eta_\alpha(t_i)$ the main quantity which enters the deconvolution
theory is the MSD function:
\begin{equation}\label{er1}
\begin{split}
\tilde\Delta_\alpha (t_i) = \langle (\tilde{z}_\alpha(t_i) - \tilde{z}_\alpha(0))^2 \rangle & = \langle ({z}_\alpha(t_i) - {z}_\alpha(0))^2 \rangle + \langle (\eta_\alpha(t_i) - \eta_\alpha(0))^2 \rangle\\
&=\Delta_\alpha (t_i) + 2\gamma,
\end{split}
\end{equation}
where $\Delta_\alpha (t_i)$ is the true MSD value.  Thus the white
noise induces a uniform upward shift of the MSD, which is irrelevant
to the theoretical analysis, since we are interested only in the MSD
slope, which determines the linear response function
$\tilde{J}_\alpha(t) = (\beta /2)d\tilde\Delta_\alpha(t)/dt =
J_\alpha(t)$.  However this assumes that we can collect an infinite
time series, reducing the statistical uncertainty of our expectation
values to zero.  In reality, our sampling is carried out only $N$
times, over a finite time interval $T_\text{s} = Nt_s$.  Our
calculated $\tilde\Delta_\alpha(t_i)$ for this time interval will
differ from the $T_\text{s} \to \infty$ result by some random amount
characterized by a standard deviation
$\sigma(\tilde\Delta_\alpha(t_i))$.  The standard error analysis for
correlation functions~\cite{Zwanzig69,Frenkel} (which assumes
Gaussian-distributed fluctuations) yields an approximate magnitude for
$\sigma(\tilde\Delta_\alpha(t_i))$, up to lowest order in $\gamma$:
\begin{equation}\label{er2}
\sigma(\tilde\Delta_\alpha(t_i)) = \sqrt{A \frac{\tau_\alpha \Gamma^2_\alpha}{N t_\text{s}} \left[1+\frac{16 t_\text{s}}{A \tau_\alpha} \frac{\gamma}{\Gamma_\alpha} + {\cal O}\left(\frac{\gamma^2}{\Gamma_\alpha^2}\right)\right]}.
\end{equation}
Here $A \sim {\cal O}(1)$ is a numerical prefactor (which may depend
on $t_i$), $\Gamma_\alpha = \langle z_\alpha^2 \rangle - \langle
z_\alpha \rangle^2$, and $\tau_\alpha = 2\Gamma_\alpha^{-2}
\int_0^\infty dt\, (\langle z_\alpha(t) z_\alpha(0)\rangle - \langle
z_\alpha\rangle^2)^2$ is an effective correlation time.  Since
$\Delta_\alpha(t)$ can be modeled as a sum of terms exponentially
converging toward the long-time limit $\Delta_\alpha(\infty) =
2\Gamma_\alpha$, the value of $\tau_\alpha$ is on the order of the
longest decay time.  Eq.~\eqref{er2} allows us to see how to correct
the effects of white noise: if in the absence of noise ($\gamma = 0$)
we could achieve a certain statistical uncertainty by taking $N_0$
steps, the same uncertainty could be obtained in the presence of white
noise by taking a larger number of steps, $N \approx N_0 (1+
16t_\text{s} \gamma/A \tau_\alpha \Gamma_\alpha)$.

Generally the larger the correlation time $\tau_\alpha$ for the
specific MSD of interest, the longer the sampling time necessary to
achieve a certain level of precision (i.e. a self MSD, involving the
motion of the individual trapped beads, will typically have a slower
relaxation time than an end-to-end MSD).  If the largest $\tau_\alpha$
is $\sim 0.1$ ms (an upper bound estimate for the dual-trap setup
discussed in the main text), $t_\text{s} = 0.01$ ms, and
$\Gamma_\alpha \sim 10$ nm$^2$ is the characteristic scale of the MSD
function, then a 1\% precision,
$\sigma(\tilde\Delta_\alpha(t_i))/\Gamma_\alpha = 0.01$, requires a
sampling time of $T_\text{s} \sim 1$ s (with no white noise).  The
same precision with a noise strength of $\gamma = 1$ nm$^2$ would need
$\sim 16\%$ more sampling time.

In contrast to the high frequency noise modeled above, imagine that we
have a low frequency contamination induced by slow drift: $\eta(t_i) =
t_i/t_\text{d} + \eta_0$ for some constants $t_\text{d}$ and $\eta_0$.
The measured MSD becomes:
\begin{equation}\label{er3}
\begin{split}
\tilde\Delta_\alpha (t_i) = \langle (\tilde{z}_\alpha(t_i) - \tilde{z}_\alpha(0))^2 \rangle & = \langle ({z}_\alpha(t_i) - {z}_\alpha(0))^2 \rangle + \langle (\eta_\alpha(t_i) - \eta_\alpha(0))^2 \rangle\\
&=\Delta_\alpha (t_i) + (t_i/t_\text{d})^2.
\end{split}
\end{equation}
Thus instead of reaching a plateau at large times, the MSD continues
to increase $\propto t^2$.  If any deviation of this kind is observed
in the experimental time series, the simplest solution is to fit the
long-time MSD, extract the functional form (i.e. the slope
$1/t_\text{d}^2$), and subtract the drift contribution from
$\tilde\Delta_\alpha (t_i)$ to recover $\Delta_\alpha (t_i)$.
Alternatively, since the drift time scale is typically larger than any
characteristic relaxation time in the system, $t_\text{d} \gg
\tau_\alpha$, one could collect data from many short runs of length
$T_\text{s}$, where $\tau_\alpha \ll T_\text{s} \ll t_\text{d}$.

\subsection{Limited time resolution}

In addition to the possibilities of noise and drift modifying the
signal, we have to consider that the measuring apparatus will have
some finite time resolution $t_\text{s}$, defined as the interval
between consecutive recordings of the data.  The measured value,
$\tilde{z}_\alpha(t_i)$, even in the absence of noise contributions
$\eta(t_i)$, only approximately corresponds to the instantaneous true
value $z_\alpha(t_i)$.  Because of the finite response time of the
equipment, it is more realistic to model $\tilde{z}_\alpha(t_i)$ as
some weighted average from the previous $t_\text{s}$ interval:
\begin{equation}\label{er4}
\tilde{z}_\alpha(t_i) = W^{-1} \int_{-t_\text{s}}^0 ds\,z_\alpha(t_i+s) w(s),\qquad W = \int_{-t_\text{s}}^0 ds\,w(s),
\end{equation}
for some weighting function $w(s)$.  In the present discussion we will
ignore any contributions from white noise and drift in the
time-averaged signal, since these can be corrected for in the same
manner as described above.  The measured autocorrelation function
$\tilde{R}_\alpha(t_i-t_j) = \langle
\tilde{z}_\alpha(t_i)\tilde{z}_\alpha(t_j) \rangle$ between two time
points $t_i$ and $t_j$ is related to the true function
$R_\alpha(t_i-t_j) = \langle
z_\alpha(t_i)z_\alpha(t_j) \rangle$ as follows:
\begin{equation}\label{er5}
\begin{split}
\tilde{R}_\alpha(t_i - t_j) &= W^{-2} \int_{-t_\text{s}}^0 ds \int_{-t_\text{s}}^0 ds^\prime \, R_\alpha(t_i + s - t_j -s^\prime) w(s) w(s^\prime)\\
&\equiv \int_{-\infty}^{\infty} dt^\prime\, B(t^\prime) R_\alpha(t_i - t_j - t^\prime),
\end{split}
\end{equation}
where we introduce the combined weighting function:
\begin{equation}\label{er6}
B(t) = \frac{1}{2W^2} \Theta(t+t_\text{s})\Theta(-t+t_\text{s}) \int_{|t|-2t_\text{s}}^{-|t|} dt^\prime\, w\left(\frac{t^\prime-t}{2}\right)w\left(\frac{t^\prime+t}{2}\right).
\end{equation}
While the precise weighting function may vary depending on the
apparatus, a reasonable approximation is that $w(s) \approx 1$, or
that the signal is directly an average over the $t_\text{s}$ interval.
Plugging this into Eq.~\eqref{er6}, with $W = t_\text{s}$, we find:
\begin{equation}\label{er7}
B(t) = \Theta(t+t_\text{s})\Theta(-t+t_\text{s}) \frac{1}{t_\text{s}}\left(1-\frac{|t|}{t_\text{s}} \right).
\end{equation}

The autocorrelation $R_\alpha(t)$ enters the deconvolution analysis
through the MSD $\Delta_\alpha(t) = 2(R_\alpha(0) - R_\alpha(t))$, and
its derivative $J_\alpha(t) = (\beta/2) d\Delta_\alpha(t)/dt = -\beta
dR_\alpha(t)/dt$.  If $R_\alpha(t)$ is expressed as a sum of decaying
exponentials, $R_\alpha(t) = \sum_i A_i \exp(-\Lambda_i t)$, then this
corresponds to $J_\alpha(t) = \sum_i C_i \exp(-\Lambda_i t)$ with the
coefficients $C_i = \beta \Lambda_i A_i$.  Assuming this underlying
exponential form, the relationship between the measured and actual
$R_\alpha(t)$ can be calculated using Eqs.~\eqref{er5} and \eqref{er7}:
\begin{equation}\label{er8}
\tilde{R}_\alpha(t) = \int_{-\infty}^{\infty} dt^\prime\, B(t^\prime) R_\alpha(t - t^\prime) = \sum_i \frac{2(\cosh(\Lambda_i t_\text{s}) -1)}{\Lambda_i^2 t_\text{s}^2} A_i e^{-\Lambda_i t}\equiv \sum_i \tilde{A}_i e^{-\Lambda_i t}, \qquad t \ge t_\text{s}.
\end{equation}
Thus the only modification is in the coefficient of each exponential
term, $\tilde A_i = p(\Lambda_i t_s) A_i$, which gets a prefactor
$p(x) = 2(\cosh(x) -1)/x^2 > 1$ dependent on $x = \Lambda_i t_\text{s}$.  The
restriction to times $t \ge t_\text{s}$ comes from the fact that
$\tilde{R}_\alpha(t)$ cannot be measured for times smaller than the
experimental resolution.  The prefactor $p(x)$ is only
appreciably larger than 1 for relaxation times $\Lambda_i^{-1}$
comparable to or smaller than $t_\text{s}$.  However, even for
$\Lambda_i^{-1} = t_\text{s}$, which is roughly the smallest
relaxation time we can realistically fit from the measured data,
$p(1) = 1.09$, so the averaging effect is modest.  

Thus when we fit a sum of exponentials to the $J_\alpha(t)$ functions
derived from the experimental time series, we are effectively
calculating $\tilde{C}_i = \beta \Lambda_i \tilde{A}_i$ and
$\Lambda_i$.  To recover the actual $C_i$, we just divide out the
prefactor: $C_i = \tilde{C}_i / p(\Lambda_i t_\text{s})$.  In this way
we correct for the distortion due to time averaging.

\begin{figure}[!t]
\centerline{\includegraphics[scale=1]{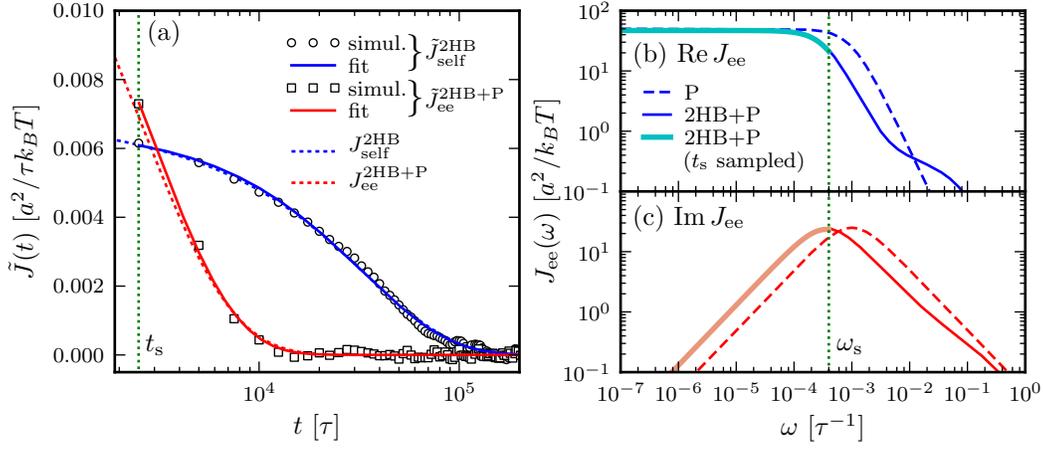}}
\caption{(a) Time-domain response functions
  $\tilde{J}_\text{self}^\text{2HB}(t)$ (circles) and
  $\tilde{J}_\text{ee}^\text{2HB+P}(t)$ (squares) calculated from
  coarse-grained simulation time series of the 2HB and 2HB+P systems
  respectively.  The coarse-graining involved averaging over intervals
  of $t_\text{s} = 2500\tau = 0.01$ ms, and using the time series of
  mean values.  The solid lines are the results of single-exponential
  fitting to the data.  The original fit results based on the fine-grained
  time series, ${J}_\text{self}^\text{2HB}(t)$ and
  ${J}_\text{ee}^\text{2HB+P}(t)$, are shown as dashed lines for
  comparison.  (b,c) The thin dashed and solid lines are the
  theoretical predictions for the Fourier-domain end-to-end responses
  $J^\text{P}_\text{ee}(\omega)$ and
  $J^\text{2HB+P}_\text{ee}(\omega)$ respectively; panel (b) shows the
  real part, panel (c) the imaginary part.  These are taken from
  Fig.~3 in the main text.  The thick solid lines are Re and Im
  $J^\text{2HB+P}_\text{ee}(\omega)$ as determined from the
  time-domain best-fit in panel (a), after correcting for the
  averaging effect.  The vertical dotted line marks the frequency
  cutoff $\omega_\text{s} = t_\text{s}^{-1}$ for the coarse-grained
  data.}\label{samp}
\end{figure}

To illustrate the effects of the experimental time resolution, and how
the above corrections would work in practice, we will redo the
deconvolution analysis for the toy protein example described in the
main text.  However, instead of using the original simulation time
series, whose data collection interval was 1.2 ns, we will average the
data over windows of width $t_\text{s} = 2500\tau = 0.01$ ms, and use
the time series of these mean values, spaced at intervals of
$t_\text{s}$ (with a total sampling time $T_\text{s} \approx 0.5$ s).
We thus roughly mimic the apparatus of Ref.~\cite{Gebhardt10}.
Fig.~\ref{samp}(a) shows time-domain response functions derived from
the coarse-grained data: $\tilde{J}_\text{self}^\text{2HB}(t)$, from
the first deconvolution step, involving the beads and handles alone;
$\tilde{J}_\text{ee}^\text{2HB+P}(t)$, from the second step, where the
toy protein is included in the system.  The fits, drawn as solid
lines, involve only single exponentials, since these are sufficient to
capture the behaviors in the restricted time range $t\ge t_\text{s}$.
From the Fourier transform of the fit results for
$\tilde{J}_\text{ee}^\text{2HB+P}(t)$ (with the $\tilde{C}_i \to C_i$
correction) we can get the true frequency space linear response of the
2HB+P system, $J_\text{ee}^\text{2HB+P}(\omega)$.  The real and
imaginary parts of this response are drawn as thick solid lines in
Fig.~\ref{samp}(b) and (c) respectively.  As expected, we can
calculate this response only up to the cutoff frequency
$\omega_\text{s} = t_\text{s}^{-1} = 4\times 10^{-4} \tau^{-1}$ of the
measuring equipment.  However, it coincides very well with the
theoretical prediction (thin solid lines, taken from Fig.~3 of the
main text).

Even though we are limited to the range $\omega < \omega_\text{s}$,
the convolution rules still work: a relation like Eq.~[3] in the main
text is valid at all individual values of $\omega$.  Though we see
only a restricted portion of the total system end-to-end response
$J_\text{ee}^\text{2HB+P}(\omega)$, it does exhibit non-trivial
structure for $\omega < \omega_\text{s}$, i.e. the downturn in the
real part and leveling off of the imaginary part at $\omega \gtrsim
10^{-4} \tau^{-1}$.  Deconvolution allows us to extract the physical
content of this structure: when numerical fitting is carried out over
the range $\omega < \omega_\text{s}$, we find toy protein parameters:
$\mu_\text{P} = 0.0472 \pm 0.005 \: \mu_0$, $k_\text{P} = 0.0201 \pm
0.0001\:k_BT/a^2$, within 6\% of the exact values ($0.05\:\mu_0$ and
$0.02\:k_BT/a^2$).

What is the main reason behind the relative success of the fitting in
extracting the protein parameters, despite the limited time
resolution?  Fig.~\ref{samp}(b,c) also shows the theoretical response
$J^\text{P}_\text{ee}(\omega)$ for the protein alone, which has a
characteristic frequency $\mu_\text{P} k_\text{P} = 10^{-3}
\tau^{-1}$, above the equipment cutoff $\omega_\text{s} = 4\times
10^{-4} \tau^{-1}$.  (Graphically, the peak in
$J^\text{P}_\text{ee}(\omega)$ falls to the right of the vertical
dotted line marking $\omega_\text{s}$.)  However, the protein dynamics
is modified by the presence of the handles and beads in the
experimental system, in a manner quantitatively described by our
convolution theory.  As a result, the characteristic frequency of the
protein within the experimental apparatus (i.e. the peak position in
the imaginary part) is shifted to $\omega = 3.6\times
10^{-4}\tau^{-1}$, just within the cutoff.  Hence we are able to fit
for the protein properties.  The relative accuracy of the fitting is
notable, since we are essentially at the very limit of resolvability.
This example illustrates a general principle: whatever dynamical
features a protein (or any other component) exhibits at $\omega <
\omega_\text{s}$ within the full experimental setup, our deconvolution
theory should be able to extract.  For this purpose, knowledge of the
system behavior for $\omega > \omega_\text{s}$ is not necessary.  In
the current case, the theoretical 2HB+P end-to-end response has a
contribution at $\omega \gtrsim 10^{-2} \tau^{-1}$ due to the DNA
handle fluctuation modes, i.e. the shoulder seen in
Fig.~\ref{samp}(b,c).  These modes are experimentally inaccessible at
our time resolution $t_\text{s}$, so the coarse-grained time series
reveals nothing about their properties.  However, this does not stop
us from applying the theory at $\omega < \omega_\text{s}$.

\end{widetext}
\end{document}